\RequirePackage{ifpdf}
\documentclass[hyper]{JHEP3}

\pdfoutput=1
\usepackage{epsfig}
\usepackage{cite}
\usepackage{graphicx}
\usepackage{subfigure}
\usepackage{inputenc}
\usepackage{amsmath}
\usepackage{snapshot}

\def\beq{\begin{equation}}
\def\eeq{\end{equation}}
\def\beeq{\begin{eqnarray}}
\def\eeeq{\end{eqnarray}}

\def\HW{\texttt{Herwig++}}
\def\PY{\texttt{Pythia8}}

\def\aMC{\texttt{aMC@NLO}}
\def\nn{\nonumber}
\def\as{\alpha_{\mathrm{S}}}
\def\aR{a_R}
\def\bas{\aR}
\def\res{{\rm res.}}
\def\ms{${\overline {\rm MS}}$}

\def\tlam{\widetilde{\lambda}}
\def\bI{\overline{\cal I}}

\def\bB{\overline B}
\def\bG{\overline G}
\def\bH{\overline H}
\def\cI{{\cal I}}
\def\cJ{{\cal J}}
\def\cO{{\cal O}}
\def\z3{\zeta_3}
\def\gE{\gamma_{\rm E}}

\title{Resummation of the transverse-energy distribution in Higgs boson production at
  the Large Hadron Collider}

\author{Massimiliano Grazzini$^a$\footnote{On leave of absence from
    INFN, Sezione di Firenze, Sesto Fiorentino, Florence, Italy.},
Andreas Papaefstathiou$^a$, Jennifer M.\
  Smillie$^b$ and Bryan R.\ Webber$^c$\\
   $^a$Physik Institut, Universit\"at Z\"urich, Switzerland\\
   $^b$Higgs Centre for Theoretical Physics, University of Edinburgh, UK\\
   $^c$Cavendish Laboratory, J.J.\ Thomson Avenue, Cambridge, UK\\
        E-mail: \email {grazzini@physik.uzh.ch},\email{andreasp@physik.uzh.ch}, \email{j.m.smillie@ed.ac.uk}, \email{webber@hep.phy.cam.ac.uk}
        }

\received{\today}               
\accepted{\today}               

\preprint{Cavendish-HEP-14/01\\Edinburgh 2014/07\\
IPMU 14-0051\\ZU-TH 09/14\\MCnet-14-06\\LPN14-052}

\abstract{
We compute the resummed hadronic transverse-energy ($E_T$)
distribution due to initial-state QCD radiation in the production of
a Standard Model Higgs boson of mass 126 GeV by gluon fusion
at the Large Hadron Collider, with
matching to next-to-leading order calculations at large $E_T$.
Effects of hadronization, underlying event
and limited detector acceptance are estimated using
\aMC\ with the \HW\ and \PY\ event generators.
}


\begin{document} 

\section{Introduction}
\label{sec:intro}
The new particle discovered recently by the ATLAS~\cite{Aad:2012tfa} and
CMS~\cite{Chatrchyan:2012ufa} Collaborations at the LHC looks very
much like the Higgs boson of the Standard Model, although its
properties remain to be fully explored.  For this exploration, detailed
predictions of the expected characteristics of Higgs production within
the Standard Model will be essential, in order to optimize signal to
background ratios and to search for any signs of new physics.  One
such characteristic is the amount and distribution of initial-state
QCD radiation, which is predicted to be exceptionally high in
production by gluon fusion and exceptionally low in vector boson
fusion.  A thorough understanding of initial-state radiation is
therefore essential for the separation of these production mechanisms.

In the present paper we study the distribution of the total amount of transverse
energy ($E_T$) emitted in Standard Model Higgs boson production by gluon fusion at the
LHC.  Results are presented at next-to-leading order (NLO) in QCD perturbation
theory and also resummed to all orders in the QCD coupling $\as$. 
The resummation applies to leading, next-to-leading and some important
next-to-next-to-leading logarithms of $E_T/Q$ ((N)NLL) where $Q$ is
the hard process scale, taken to be the Higgs mass $m_H$.
Thus it improves the treatment of the small-$E_T$ region, where the fixed-order
prediction diverges whereas the actual distribution must tend to zero as $E_T\to
0$.  By matching the resummed prediction to the NLO result valid at
large $E_T$, we provide a uniform description from the low-$E_T$ to the
high-$E_T$ region.

Our approach follows on from Ref.~\cite{Papaefstathiou:2010bw}, based
in turn on the early work on $E_T$ resummation in vector-boson
production~\cite{Halzen:1982cb,Davies:1983di,Altarelli:1986zk} and
closely related to the resummation of transverse momentum in
vector-boson~\cite{Parisi:1979se,Collins:1984kg,Bozzi:2010xn,Becher:2010tm}
and Higgs
production~\cite{Bozzi:2005wk,Bozzi:2007pn,Mantry:2009qz,Catani:2010pd}.\footnote{The resummation of the jet-veto $p_T$ distribution has been considered in
  Refs.~\cite{Banfi:2012yh,Becher:2012qa,Banfi:2012jm,Becher:2013xia,Stewart:2013faa}.}
We make a number of improvements relative to Ref.~\cite{Papaefstathiou:2010bw},
including:
\begin{itemize}
\item Predictions for the experimentally relevant Higgs mass of 126 GeV, at
centre-of-mass energies $\sqrt{s}=8$ and $14$ TeV;
\item Fixed-order predictions to NLO, i.e.\ $\cO(\as^4)$.
\item Expansion of the $E_T$ resummation formula to NLO, and
  demonstration that to this order the structure of the logarithmic terms
  is consistent with the fixed-order prediction;
\item Matching of the resummed and NLO predictions across the whole
  range of $E_T$;
\item A constraint on the perturbative unitarity of the prediction, using the
  method of Ref.~\cite{Bozzi:2005wk}, which reduces the impact of
  logarithmic terms in the large-$E_T$ region;
\item Studies of the effects of renormalization scale variation and
  unknown higher-order terms;
\item Monte Carlo studies of the effects of hadron-level cuts on
  pseudorapidity and transverse momentum, with fixed-order matching to 
parton showers using \aMC\ interfaced to \HW\ and \PY.
\end{itemize} 

The paper is organized as follows.  In Sec.~2, we review the resummation
procedure and then describe the necessary modifications to implement the
unitarity condition mentioned above.  This involves some changes in
the formalism and the
evaluation of new integrals in this prescription.  In Sec.~3, we expand our
resummed result to next-to-leading order in order to match our results to the
fixed-order prediction at this accuracy.  This renders our predictions positive
throughout the $E_T$-range.  In Sec.~4, we investigate the $E_T$ distribution
further through Monte Carlo studies.  We first reweight Monte Carlo results to
our analytic distribution and then investigate the impact of hadronisation and
underlying event.  We end the main text in Sec.~5 with conclusions and
discussion.  A number of appendices then contain supplementary results.

\section{Resummation of logarithmically enhanced terms}\label{sec:method}
Here we summarize the results of
Ref.~\cite{Papaefstathiou:2010bw} as applied to Higgs boson production.
The resummed component of the transverse-energy distribution in the
process $h_1h_2\to HX$ at scale $Q$ has the form
\beeq
\label{resgen}
\left[ \frac{d\sigma_H}{dQ^2\;dE_T} \right]_{\res} &=& \frac 1{2\pi}\sum_{a,b}
\int_0^1 dx_1 \int_0^1 dx_2 \int_{-\infty}^{+\infty} d\tau \; {\rm e}^{-i\tau
  E_T}
\;f_{a/h_1}(x_1,\mu) \; f_{b/h_2}(x_2,\mu) \nn \\
&\cdot& W_{ab}^H(x_1 x_2 s; Q, \tau,\mu) \eeeq where $f_{a/h}(x,\mu)$ is the
parton distribution function (PDF) of parton $a$ in hadron $h$ at factorization
scale $\mu$, taken to be the same as the renormalization scale here (we
illustrate the impact of varying this scale in Sec.~3).  In what follows we use
the \ms\ renormalization scheme.  To take into account the constraint that the
transverse energies of the emitted partons should sum to $E_T$, the resummation
procedure is carried out in the domain that is Fourier conjugate to $E_T$.  The
transverse-energy distribution (\ref{resgen}) is thus obtained by performing the
inverse Fourier transformation with respect to the ``transverse time'', $\tau$.
The factor $W_{ab}^H$ is the perturbative and process-dependent partonic cross
section that embodies the all-order resummation of the large logarithms $\ln
(Q\tau)$.  Since $\tau$ is conjugate to $E_T$, the limit $E_T\ll Q$ corresponds
to $Q\tau \gg 1$.

As in the case of transverse-momentum resummation~\cite{Catani:2000vq},
the resummed partonic cross section can be written in the following form:
\beeq
\label{eq:Wab}
W_{ab}^H(s; Q, \tau,\mu) &=& \int_0^1 dz_1 \int_0^1 dz_2 
\; C_{ga}(\as(\mu), z_1;\tau,\mu) \; C_{gb}(\as(\mu), z_2;\tau,\mu)
\; \delta(Q^2 - z_1 z_2 s) \nn\\
&\cdot& \sigma_{gg}^H(Q,\as(Q)) \;S_g(Q,\tau) \;\;.
\eeeq
Here $\sigma_{gg}^H$ is the cross section
for the partonic subprocess of
 gluon
fusion, $gg \to H$, through a massive-quark loop:
\beq\label{eq:higgsLO}
\sigma_{gg}^H(Q,\as(Q)) = \delta(Q^2-m_H^2)\sigma_0^H\;,
\eeq
where in the limit of infinite quark mass
\beq
\sigma_0^H = \frac{\as^2(m_H)G_Fm_H^2}{288\pi\sqrt{2}}\;.
\eeq
$S_g(Q,\tau)$ is the appropriate
gluon form factor, which in the case of $E_T$ resummation
takes the form~\cite{Davies:1983di,Altarelli:1986zk}
\beq
\label{formfact}
S_g(Q,\tau) = \exp \left\{-2\int_0^Q \frac{dq}q 
\left[ 2A_g(\as(q)) \;\ln \frac{Q}{q} + B_g(\as(q)) \right] 
\left(1-{\rm e}^{iq\tau}\right)\right\} \;.
\eeq
The functions $A_g(\as), B_g(\as)$, as well as the 
coefficient functions $C_{ga}$ in Eq.~(\ref{eq:Wab}), contain no
$\ln (Q\tau)$ terms and are perturbatively computable as power
expansions with constant coefficients:
\beeq
\label{aexp}
A_g(\as) &=& \sum_{n=1}^\infty \left( \frac{\as}{\pi} \right)^n A_g^{(n)} 
\;\;, \\
\label{bexp}
B_g(\as) &= &\sum_{n=1}^\infty \left( \frac{\as}{\pi} \right)^n B_g^{(n)}
\;\;, \\
\label{cexp}
C_{ga}(\as,z) &=& \delta_{ga} \,\delta(1-z) + 
\sum_{n=1}^\infty \left( \frac{\as}{\pi} \right)^n C_{ga}^{(n)}(z) \;\;.
\eeeq
Thus a calculation to NLO in $\as$ involves the coefficients
$A_g^{(1)}$,  $A_g^{(2)}$,  $B_g^{(1)}$,  $B_g^{(2)}$ and $C_{ga}^{(1)}$.
The coefficients $A_g^{(1)}$, $A_g^{(2)}$, $B_g^{(1)}$ and
$C_{ga}^{(1)}$ read~\cite{Catani:1988vd,Kauffman:1992cx}
\beeq\label{eq:Agetc} &&A_g^{(1)} = C_A \;,\quad A_g^{(2)} = \frac{1}{6} C_A
\left[ C_A\left(\frac{67}6-\frac{\pi^2}2\right)-\frac 53 n_f\right] \;,\quad
B_g^{(1)} = - \frac{1}{6} (11 C_A - 2 n_f) \;,\nn\\
&&C_{gg}^{(1)}(z)=\frac 14
\left[C_A\left(2-\frac{\pi^2}3\right)+5+4\pi^2\right]\delta(1-z)
\equiv c^{(1)}_g\delta(1-z)\;,\nn\\
&&C_{gq}^{(1)}(z)=C_{g\bar q}^{(1)}(z)=\frac 12 C_F z\,.  \eeeq The coefficient
$B_g^{(2)}$ for the Higgs transverse-momentum spectrum
is~\cite{deFlorian:2000pr,deFlorian:2001zd}
\beq\label{eq:B2g} \bB_g^{(2)}=C_A^2\left(\frac{23}{24}+
  \frac{11}{18}\pi^2-\frac{3}{2}\z3\right) +\frac{1}{2}
C_F\,n_f-C_A\,n_f\left(\frac{1}{12}+\frac{\pi^2}{9} \right) -\frac{11}{8} C_F
C_A\, .  \eeq However, the value of the coefficient $B_g^{(2)}$ for the
transverse energy in Higgs production could be
different\footnote{We are grateful to Jon Walsh for a useful
  discussion on this point.}.  In Sec.~3, we will perform a fit to
the fixed-order NLO result at small transverse energy, with this
coefficient as a free parameter.

Returning to Eq.~(\ref{resgen}), we may recast it in a form with a real
integrand as
\beeq\label{eq:resF}
\left[ \frac{d\sigma_H}{dQ^2\;dE_T} \right]_{\res} &=&
\frac 1{\pi s}\int_{0}^{\infty}
 d\tau \; {\rm e}^{-F_g^{(R)}(Q,\tau)} \Bigl[
 R_g^{(R)}(s;Q,\tau)\cos\{F_g^{(I)}(Q,\tau)+\tau E_T\}\nn\\
&&+R_g^{(I)}(s;Q,\tau)\sin\{F_g^{(I)}(Q,\tau)+\tau E_T\}\Bigr]
\,\sigma_{gg}^H(Q,\as(Q))
\eeeq
where $F_g^{(R)}$ and $F_g^{(I)}$ are the real and imaginary parts of
\beq\label{eq:Fcs}
F_g(Q,\tau) = 2\int_0^Q \frac{dq}q 
\left[ 2A_g(\as(q)) \;\ln \frac{Q}{q} + B_g(\as(q)) \right] 
\left(1-{\rm e}^{iq\tau}\right)\,.
\eeq

As explained in~\cite{Papaefstathiou:2010bw}, the coefficient
functions in Eq.~(\ref{eq:Wab}) contain logarithms of $\mu\tau$,
which are eliminated by the choice of factorization
scale $\mu_F=\tau_0/\tau$, where $\tau_0 = \exp(-\gE)=0.56146\ldots$,
$\gE$ being the Euler-Mascheroni constant.
The resulting expressions for $R_g^{(R,I)}$ are
\beeq\label{eq:RgRI}
&R_g^{(R)}(\xi,\tau)&=
\int_{\xi}^1 \frac{dx_1}{x_1}\Biggl\{f_{g/h_1}(x_1)f_{g/h_2}\left(
\frac{\xi}{x_1}\right)\left(1+\frac{\as}{\pi}2c^{(1)}_g\right)\nn\\
&& + \frac{\as}{\pi}\int_{\xi/x_1}^1\frac{dz}z
\left[f_{g/h_1}(x_1)f_{s/h_2}\left(\frac{\xi}{zx_1}\right)+ f_{s/h_1}(x_1)
f_{g/h_2}\left(\frac{\xi}{zx_1}\right)\right]\frac 12 C_F z\Biggr\}\;,\nn\\
&R_g^{(I)}(\xi,\tau)&=
\frac{\as}2\int_{\xi}^1 \frac{dx_1}{x_1} \int_0^1\frac{dz}z
\Biggl\{2f_{g/h_1}(x_1)f_{g/h_2}\left(\frac{\xi}{zx_1}\right)P_{gg}(z)\nn\\
&&+\left[f_{g/h_1}(x_1)f_{s/h_2}\left(\frac{\xi}{zx_1}\right)
+f_{s/h_1}(x_1)f_{g/h_2}\left(\frac{\xi}{zx_1}\right)\right]P_{gq}(z)\Biggr\}\nn\\
&=&\hspace{-0.3cm}\frac{\as}2\int_{\xi}^1 \frac{dx_1}{x_1}\Biggl\{
2f_{g/h_1}(x_1)f_{g/h_2}\left(\frac{\xi}{x_1}\right)
\left[2C_A\ln\left(1-\frac{\xi}{x_1}\right)+\frac 16 (11C_A-2n_f)\right]\nn\\
&&\hspace{-1cm}+\int_{\xi/x_1}^1\hspace{-0.2cm}\frac{dz}z\Biggl[4C_Af_{g/h_1}(x_1)\left\{f_{g/h_2}\left(\frac{\xi}{zx_1}\right)
\left[\frac{z}{1-z}+\frac{1-z}z+z(1-z)\right]
- f_{g/h_2}\left(\frac{\xi}{x_1}\right)\frac{z}{1-z}\right\} \nn\\
&&\hspace{-1cm}+\left\{f_{g/h_1}(x_1)f_{s/h_2}\left(\frac{\xi}{zx_1}\right)
+f_{s/h_1}(x_1)f_{g/h_2}\left(\frac{\xi}{zx_1}\right)\right\}C_F\frac{1+(1-z)^2}z\Biggr]\Biggr\}, 
\eeeq
where $f_s=\sum_q (f_q+f_{\bar q})$ and all PDFs and coefficient functions are
understood to be evaluated at scale $\mu_F=\tau_0/\tau$.  We have defined
$\xi=Q^2/s$ for convenience.

\subsection{Evaluation of the exponent}
We now seek to evaluate the exponent of the form factor,
(\ref{eq:Fcs}), analytically.  We will
use the method of Ref.~\cite{Bozzi:2005wk} where the analogous calculation was
performed for transverse-momentum resummation.  In the notation of that paper,
we have, for a renormalization scale $\mu_R$,
\beq
\label{Gformfact}
G_g(\bas,L) \equiv -2\int_{b_0/b}^Q \frac{dq}q 
\left[ 2A_g(\as(q)) \;\ln \frac{Q}{q} + B_g(\as(q)) \right] 
=L\,g_1(Y)+g_2(Y)+\bas\, g_3(Y)+\ldots
\eeq
where $\bas=\as(\mu_R)/\pi$,  $L=2\ln(Qb/b_0)$, $Y=\beta_0 \bas L$ and
$\beta_0 = (11C_A-2n_f)/12$ is the lowest-order coefficients of the beta function:
\beq
\frac{d \ln \bas}{d \ln \mu_R^2} = \beta(\bas) = - \sum_{n=0}^\infty
\beta_n \bas^{n+1}.
\eeq
The term $Lg_1(Y)$ collects the LL contributions $\as^n L^{n+1}$, the
function $g_2$ resums the NLL contributions $\as^n L^n$, the function
$g_3$ controls the NNLL terms $\as^nL^{n-1}$, and so forth.  We will
give the explicit form of the $g_i$ functions below.  We can therefore
deduce that in general
\beq\label{eq:Q0Qint}
-2\int_{Q_0}^Q \frac{dq}q 
\left[ 2A_g(\as(q)) \;\ln \frac{Q}{q} + B_g(\as(q)) \right] 
=2\lambda g_1(y)+g_2(y)+\aR\, g_3(y)+\ldots
\eeq
where now 
\beq\label{eq:lamdef}
y\equiv 2\beta_0\bas\lambda \qquad {\rm and} \qquad \lambda \equiv
\ln(Q/Q_0)
 \;.
\eeq
Now by expressing $\as(q)$ in terms of $\as(Q)$ and relating this in
turn to $\as(\mu_R)$, we can write the integrand in
(\ref{eq:Q0Qint}) as a function of $\ln(Q/q)$, and then use the result
\beq\label{eq:logints}
\int_{Q_0}^Q \frac{dq}q f\left(\ln\frac Qq\right)
=f\left(\frac d{du}\right)\frac 1u\left.\left({\rm e}^{\lambda u}-1\right)\right|_{u=0}\,,
\eeq
which is easily seen by expanding the exponential.
This allows one to calculate the $g_i$ functions explicitly.  They are given
by~\cite{Frixione:1998dw,deFlorian:2004mp,Bozzi:2005wk}:
\beeq \label{eq:gs}
g_1(y) &=& \frac{A_g^{(1)}}{\beta_0 y} \left( y+\ln(1-y) \right)\,, \nonumber \\
g_2(y) &=& \frac{B_g^{(1)}}{\beta_0} \ln(1-y) - \frac{A_g^{(2)}}{\beta_0^2}
\left( \frac{y}{1-y} + \ln(1-y) \right) \nonumber \\ 
 && \  + \frac{A_g^{(1)} \beta_1}{\beta_0^3}
\left( \frac12 \ln^2(1-y) + \frac{y+\ln(1-y)}{1-y} \right)  +
\frac{A_g^{(1)}}{\beta_0} \left( \frac{y}{1-y}+\ln(1-y) \right) \ln 
 \left( \frac{Q^2}{\mu_R^2} \right) \,, \nonumber\\
g_3(y) &=&-\frac{A_g^{(3)}}{2\beta_0^2}\left(\frac y{1-y}\right)^2
  -\frac{B_g^{(2)}}{\beta_0} \frac{y}{1-y} + \frac{A_g^{(2)}
  \beta_1}{\beta_0^3} \frac{y(3y-2)-2(1-2y)\ln(1-y)}{2(1-y)^2}\\ 
  && \ +\frac{A_g^{(1)}}{\beta_0^4} \left( \frac{\beta_1^2
      (1-2y)\ln^2(1-y)}{2(1-y)^2} + \ln(1-y) \left( \beta_0\beta_2-\beta_1^2 +
      \frac{\beta_1^2}{1-y} \right) \right. \nonumber \\
  && \qquad \left. + \frac{y(\beta_0 \beta_2 (2-3y)+ \beta_1^2
      y)}{2(1-y)^2} \right) + \frac{B_g^{(1)}
  \beta_1}{\beta_0^2} \frac{y+\ln(1-y)}{1-y} - \frac{A_g^{(1)}}2
\frac{y^2}{(1-y)^2} \ln^2\left( \frac{Q^2}{\mu_R^2} \right)\nonumber \\ 
  && \ +\left( B_g^{(1)} \frac{y}{1-y} + \frac{A_g^{(2)}}{\beta_0} \frac{y^2}{(1-y)^2} +
    A_g^{(1)} \frac{\beta_1}{\beta_0^2} \left( \frac{y}{1-y} +
      \frac{(1-2y)\ln(1-y)}{(1-y)^2} \right) \right)
      \ln\left( \frac{Q^2}{\mu_R^2} \right) .\nonumber
\eeeq
Now, the actual integral we require is 
\beq\label{eq:Fint}
F_g(\as,\lambda)\equiv 2\int_0^Q \frac{dq}q 
\left[ 2A_g(\as(q)) \;\ln \frac{Q}{q} + B_g(\as(q)) \right] 
\left(1-{\rm e}^{iq\tau}\right)
\eeq
and so we must introduce $\left(1-{\rm e}^{iq\tau}\right)$ in the integrand.
The analogue of Eq.~(\ref{eq:logints}) is
\beq
\int_0^Q \frac{dq}q f\left(\ln\frac Qq\right)\left(1-{\rm e}^{iq\tau}\right)
=f\left(\frac d{du}\right)\left.\cJ(Q\tau;-u)\right|_{u=0}
\eeq
where the generating function
\beeq\label{eq:cJu}
\cJ(Q\tau;u)&=&\int_0^Q \frac{dq}q\left(\frac qQ\right)^u \left(1-{\rm
    e}^{iq\tau}\right)\nn\\
&=&\frac 1u - \left(-iQ\tau\right)^{-u}\gamma(u ,-iQ\tau)\,,
\eeeq
$\gamma(u,z)$ being the incomplete gamma function,
\beq\label{eq:gamma}
\gamma(u,z) = \Gamma(u) -z^{u-1}{\rm e}^{-z}\sum_{k=0}^\infty
\frac{\Gamma(u)}{\Gamma(u-k)}z^{-k}\;.
\eeq
The series represents power corrections, which we do not wish to
include in the resummation, so we write instead
\beeq\label{eq:cJulogs}
\cJ(Q\tau;-u)&=&\frac 1u \left[\left(-iQ\tau\right)^{u}\Gamma(1-u)-1\right]\nn\\
&=&\frac 1u \left[\exp\left(\lambda u+\sum_{k=2}\frac{\zeta_k}k
    u^k\right)-1\right]
\equiv\frac 1u \left[{\rm e}^{\lambda u}Z(u)-1\right]
\eeeq
where now
\beq
\lambda=\ln\left(\frac{Q\tau}{i\tau_0}\right)\,,
\eeq
i.e.\ we have chosen $Q_0=i{\rm e}^{-\gE}/\tau = i\tau_0/\tau$ in
(\ref{eq:lamdef}), and
\beq
Z(u)\equiv\exp\left(\sum_{k=2}\frac{\zeta_k}k
    u^k\right)=\tau_0^u\Gamma(1-u)\,.
\eeq
Now
\beq
\frac 1u \left[{\rm e}^{\lambda u}Z(u)-1\right]=
\frac{Z(u)}u \left[{\rm e}^{\lambda u}-1\right]+
\frac 1u \left[Z(u)-1\right]
\eeq
and the second term involves no logarithms, so again we drop it from the
resummation.  We show in Appendix~\ref{app:proof} that the first term
implies that
\beq\label{eq:FZG}
F_g(\as,\lambda)\equiv 2\int_0^Q \frac{dq}q 
\left[ 2A_g(\as(q)) \;\ln \frac{Q}{q} + B_g(\as(q)) \right] 
\left(1-{\rm e}^{iq\tau}\right)=
-Z\left(\frac d{d\lambda}\right)G_g(\as,2\lambda) \,,
\eeq
where $G_g$ was defined in Eq.~(\ref{Gformfact}). Now 
\beq\label{eq:Zddl}
Z\left(\frac d{d\lambda}\right)=1+\frac{\zeta_2}{2}
\frac{d^2}{d\lambda^2}+\frac{\z3}3\frac{d^3}{d\lambda^3}+\ldots
\eeq
where $\zeta_2=\pi^2/6$, so
\beq\label{eq:Fal}
F_g(\as,\lambda)
=-2\lambda g_1(y)-g_2(y) -\aR\widetilde g_3(y)+\ldots
\eeq
where
\beeq
\widetilde g_3(y)&=&g_3(y)+
\frac{\pi^2}{12\aR}\frac{d^2}{d\lambda^2}[2\lambda g_1(y)] \nonumber \\ 
&=& g_3(y)-\frac{\pi^2}3 \frac{A_g^{(1)}}{(1-y)^2}\,.
\eeeq
The other terms from
(\ref{eq:Zddl}) contribute logarithms only at the level of $g_4$ and
beyond, so we do not consider them.

Following Ref.~\cite{Bozzi:2005wk}, we can now enforce the `unitarity'
condition, $F_g\to 0$ as $\tau\to 0$, by a shift of argument of the logarithm:
\beq\label{eq:ltotl} 
\lambda\to\tlam = \ln\left(1+\frac{Q\tau}{i\tau_0}\right)
=\frac 12\ln\left(1+\frac{Q^2\tau^2}{\tau_0^2}\right)
-i\arctan\left(\frac{Q\tau}{\tau_0}\right)\,,
\eeq
so that now $y=2\beta_0\as(\mu_R^2)\tlam/\pi$.   We must apply
a corresponding shift in the factorization scale of the parton
distributions and coefficient functions (given explicitly below in
Eqs.~(\ref{eq:RgNLO})).  They are now evaluated at a scale of
\beq\label{eq:tmuF}
\widetilde\mu_F = \frac Q{\sqrt{1+Q^2\tau^2/\tau_0^2}}
\eeq
instead of $\mu_F=\tau_0/\tau$, and one must also replace
$\as(\mu_F)/2$ in the coefficient of $R_g^{(I)}$ by
\beq
\frac{\as(\widetilde\mu_F)}{\pi}\arctan\left(\frac{Q\tau}{\tau_0}\right)\,.
\eeq

We show in Appendix~\ref{app:disp} that the vanishing of the
transverse-energy distribution for $E_T<0$ implies a dispersion
relation between the real and imaginary parts of its Fourier
transform.   This allows (\ref{eq:resF}) to be written in the
simpler equivalent form
\beeq\label{eq:resFnew}
\left[ \frac{d\sigma_H}{dQ^2\;dE_T} \right]_{\res} &=&
\frac 2{\pi s}\int_{0}^{\infty}
 d\tau \; {\rm e}^{-F_g^{(R)}(Q,\tau)} \cos(\tau E_T)\Bigl[
 R_g^{(R)}(s;Q,\tau)\cos\{F_g^{(I)}(Q,\tau)\}\nn\\
&&-R_g^{(I)}(s;Q,\tau)\sin\{F_g^{(I)}(Q,\tau)\}\Bigr]
\,\sigma_{gg}^H(Q,\as(Q))
\eeeq
and implies that
\beq\label{eq:unit}
\int_0^\infty dE_T\,\left[ \frac{d\sigma_H}{dQ^2\;dE_T} \right]_{\res} =
\frac 1s R_g^{(R)}(s;Q,0)\,\sigma_{gg}^H(Q,\as(Q))\,,
\eeq
where, on account of (\ref{eq:tmuF}), the parton distributions in
$R_g^{(R)}(s;Q,0)$ are evaluated at scale $Q$.

\section{Matching to fixed order}\label{sec:match}

We now match the resummed expression derived above to the NLO
perturbative expansion of the transverse energy distribution, taking
care to  avoid double counting of the terms already contained in the
resummation.

\subsection{Expansion of the resummed prediction}
Performing the expansion of Eq.~(\ref{eq:Fal}) in powers of
$\bas\equiv\as(\mu_R^2)/\pi$, we find
\beeq
-2\lambda g_1 &=& 2 A_g^{(1)}\lambda^2 \bas + \frac 83 \beta_0 A_g^{(1)} \lambda^3\bas^2
+\cO(\bas^3)\nn\\
-g_2 &=& 2 B_g^{(1)}\lambda \bas + 2\left[A_g^{(2)}+\beta_0 B_g^{(1)}-\beta_0 A_g^{(1)}
\ln\left(\frac{Q^2}{\mu_R^2}\right)\right] \lambda^2\bas^2
+\cO(\bas^3)\nn\\
-\bas\widetilde g_3 &=& \frac{\pi^2}3 A_g^{(1)}\bas +2 \left[B_g^{(2)}+\frac 23\pi^2 \beta_0 A_g^{(1)}-\beta_0 B_g^{(1)}
\ln\left(\frac{Q^2}{\mu_R^2}\right)\right]\lambda\bas^2
+\cO(\bas^3)\,,
\eeeq
so that to NLO
\beq
F_g(\as,\lambda) =\bas F_1+\bas^2 F_2
\eeq
where, following the shift $\lambda\to\tlam$ according to Eq.~(\ref{eq:ltotl})
\beeq
F_1 &=&  2 A_g^{(1)}\left(\tlam^2 +\frac{\pi^2}6\right) + 2 B_g^{(1)}\tlam\nn\\
F_2 &=&   \frac 83 \beta_0 A_g^{(1)} \tlam^3 + 2\left[A_g^{(2)}+\beta_0 B_g^{(1)}-\beta_0 A_g^{(1)}
\ln\left(\frac{Q^2}{\mu_R^2}\right)\right] \tlam^2\nn\\
&& + 2 \left[B_g^{(2)}+\frac 23\pi^2 \beta_0 A_g^{(1)}-\beta_0 B_g^{(1)}
\ln\left(\frac{Q^2}{\mu_R^2}\right)\right]\tlam\,.
\eeeq
Similarly, evaluating all PDFs at scale $\mu=Q$, we can write to NLO
\beq
R_g(\tau) = R_0 + \bas \left(R_1 +\tlam  R'_1\right)
+ \bas^2 \left(R_2 +\tlam  R'_2+\tlam^2  R''_2\right)
\eeq
so that
\beeq
S_g R_g &=& R_0+\bas R_1 +\bas^2 R_2 +\bas\left(\tlam  R'_1-F_1R_0\right)\nn\\
&+&\bas^2\left[\tlam R'_2+\tlam^2 R''_2-(F_2-\frac 12F_1^2)R_0-F_1(R_1+\tlam  R'_1)\right]
\eeeq
where
\beeq\label{eq:RgNLO}
R_0&=&\int_{\xi}^1
\frac{dx_1}{x_1}f_g(x_1)f_g\left(\frac{\xi}{x_1}\right)\nn\\
R_1&=&\int_{\xi}^1
\frac{dx_1}{x_1}\Biggl\{2c^{(1)}_g f_g(x_1)f_g\left(\frac{\xi}{x_1}\right)
+ C_F\int_{\xi/x_1}^1 dz\, f_s(x_1)f_g\left(\frac{\xi}{zx_1}\right)\Biggr\}\nn\\
R'_1&=&-\int_{\xi}^1 \frac{dx_1}{x_1}\Biggl\{
2f_g(x_1)f_g\left(\frac{\xi}{x_1}\right)
\left[2C_A\ln\left(1-\frac{\xi}{x_1}\right)+\frac 16 (11C_A-2n_f)\right]\nn\\
&&+\int_{\xi/x_1}^1\frac{dz}z
\Biggl[4C_Af_g(x_1)\left\{f_g\left(\frac{\xi}{zx_1}\right)
\left[\frac{z}{1-z}+\frac{1-z}z+z(1-z)\right]
- f_g\left(\frac{\xi}{x_1}\right)\frac{z}{1-z}\right\} \nn\\
&&+2C_Ff_s(x_1)f_g\left(\frac{\xi}{zx_1}\right)\frac{1+(1-z)^2}z\Biggr]\Biggr\}.
\eeeq

Performing the Fourier transformation (\ref{resgen}),
we find terms involving the integrals
\beq\label{eq:Ip}
\cI_p(E_T,Q) = \frac 1{2\pi}\int_{-\infty}^{+\infty} d\tau \; {\rm e}^{-i\tau E_T} 
\ln^p\left(1+\frac{Q\tau}{i\tau_0}\right)
\eeq
with $p=1,2,3,4$, which may be evaluated from
\beq\label{eq:IpIu}
\cI_p(E_T,Q) = \frac{d^p}{du^p}\cI(E_T,Q;u)|_{u=0}
\eeq
with generating function
\beq\label{eq:Iu}
\cI(E_T,Q;u) = \frac 1{2\pi}\int_{-\infty}^{+\infty} d\tau \; {\rm e}^{-i\tau E_T}
\left(1+\frac{Q\tau}{i\tau_0}\right)^u\;.
\eeq
Writing
\beq
1+\frac{Q\tau}{i\tau_0} = \frac{zQ}{\tau_0 E_T}
\eeq
we have
\beq
\cI(E_T,Q;u) = -\frac i{2\pi E_T}\left(\frac Q{E_T\tau_0}\right)^u
\int_{-i\infty}^{+i\infty} dz\; z^u\,{\rm e}^{z-\tau_0 E_T/Q}\;.
\eeq
We can safely deform the integration contour around the branch cut along the negative real axis to obtain
\beeq\label{eq:Iufin}
\cI(E_T,Q;u) &=& -\frac u{E_T}\left(\frac Q{E_T}\right)^u
\frac{\exp(u\gE-\tau_0 E_T/Q)}{\Gamma(1-u)}\nn\\
&=&  -\frac u{E_T}\left(\frac
  Q{E_T}\right)^u\exp\left(-\frac{\tau_0 E_T}Q-\sum_{k=2}^\infty \frac{\zeta_k}k
  u^k\right)\;.
\eeeq
This gives
\beeq\label{eq:I12}
\cI_1(E_T,Q) &=& -\frac 1{E_T}{\rm e}^{-\tau_0 E_T/Q}\nn\\
\cI_2(E_T,Q) &=& -\frac 2{E_T}\ln\left(\frac Q{E_T}\right){\rm e}^{-\tau_0 E_T/Q}\nn\\
\cI_3(E_T,Q) &=& -\frac3{E_T}\left[\ln^2\left(\frac Q{E_T}\right) -\frac{\pi^2}6\right]{\rm e}^{-\tau_0 E_T/Q}\nn\\
\cI_4(E_T,Q) &=& -\frac4{E_T}\left[\ln^3\left(\frac Q{E_T}\right) -\frac{\pi^2}2\ln\left(\frac Q{E_T}\right)-2\z3\right]{\rm e}^{-\tau_0 E_T/Q}\;.
\eeeq
Therefore the NLO expansion of the resummed
expression is
\beeq\label{eq:ETNLO}
\left[\frac{E_T}{\sigma_0^H}\frac{d\sigma_H}{dE_T}\right]_{\rm resum,NLO} &=&
  \Bigl[\bas(G_0 R_0+G'_1 R'_1)+\bas^2(H_0 R_0+H_1 R_1\nn\\
&&+H'_1 R'_1+H'_2 R'_2+H''_2 R''_2)\Bigr]{\rm e}^{-\tau_0 E_T/Q}
\eeeq
where, writing $L=\ln(Q/E_T)$, 
\beeq\label{eq:GH}
G_0&=&H_1=4A_g^{(1)}L+2B_g^{(1)}\;,\;\;\; G'_1=-1\;,\nn\\
H_0&=&4L\left[A_g^{(2)}+\beta_0 B_g^{(1)} -\beta_0 A_g^{(1)}
\ln\left(\frac{Q^2}{\mu_R^2}\right)\right] 
+8\beta_0 A_g^{(1)}\left(L^2-\frac{\pi^2}6\right)\nn\\
&+&2\left[B_g^{(2)}+\frac 23\pi^2 \beta_0 A_g^{(1)}-\beta_0 B_g^{(1)}
\ln\left(\frac{Q^2}{\mu_R^2}\right)\right]
-8(A_g^{(1)})^2\left(L^3-\frac{\pi^2}3L-2\z3\right)\nn\\
&-&12A_g^{(1)}B_g^{(1)}\left(L^2-\frac{\pi^2}9\right)-4(B_g^{(1)})^2 L\;,\nn\\
H'_1&=&6A_g^{(1)}\left(L^2-\frac{\pi^2}9\right)+4B_g^{(1)} L\;, \;\;\;
H'_2 = -1\;,\;\;\; H''_2=-2L\;.
\eeeq

\begin{figure}
\begin{center}
\epsfig{file=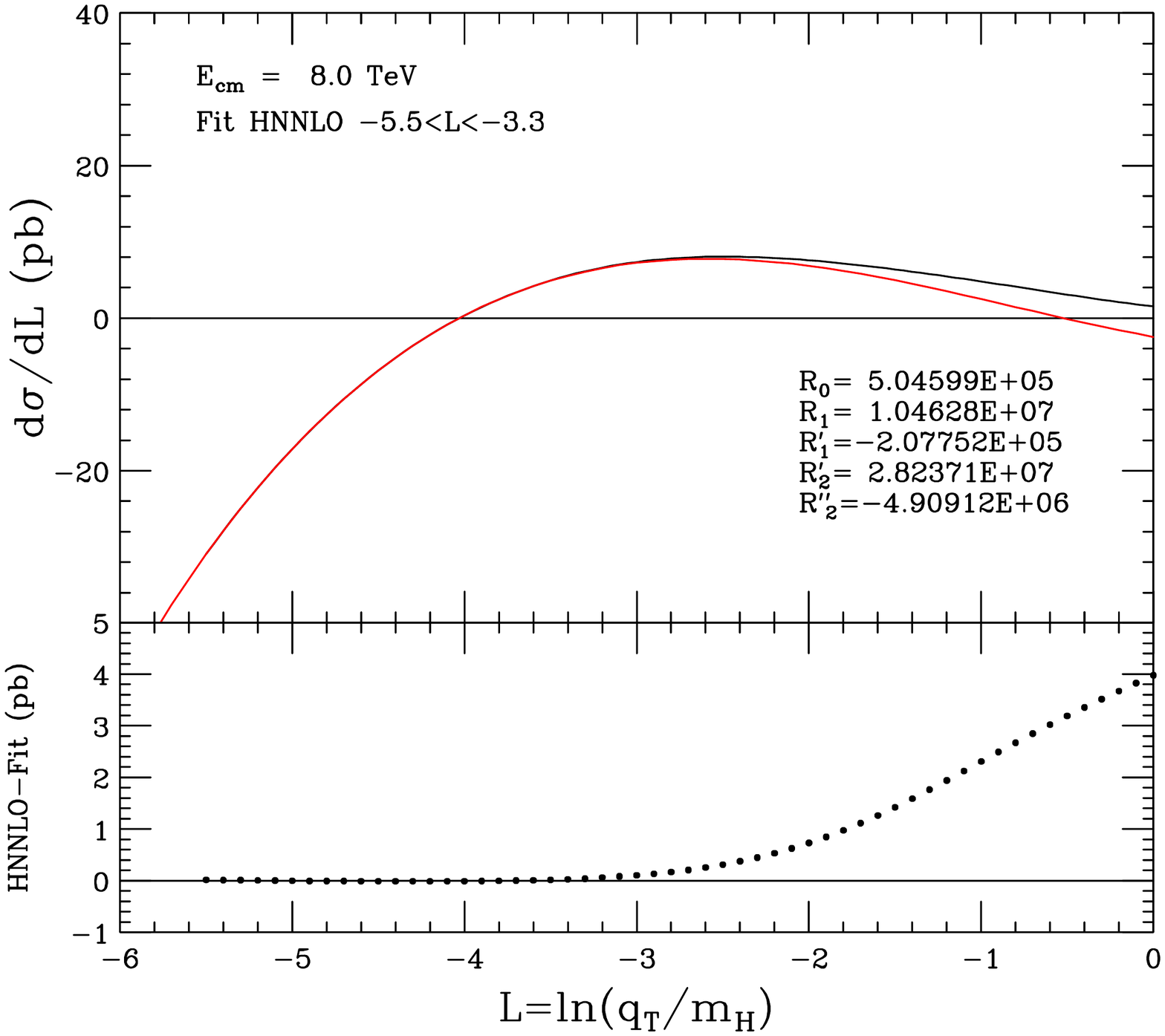,width=75mm}
\epsfig{file=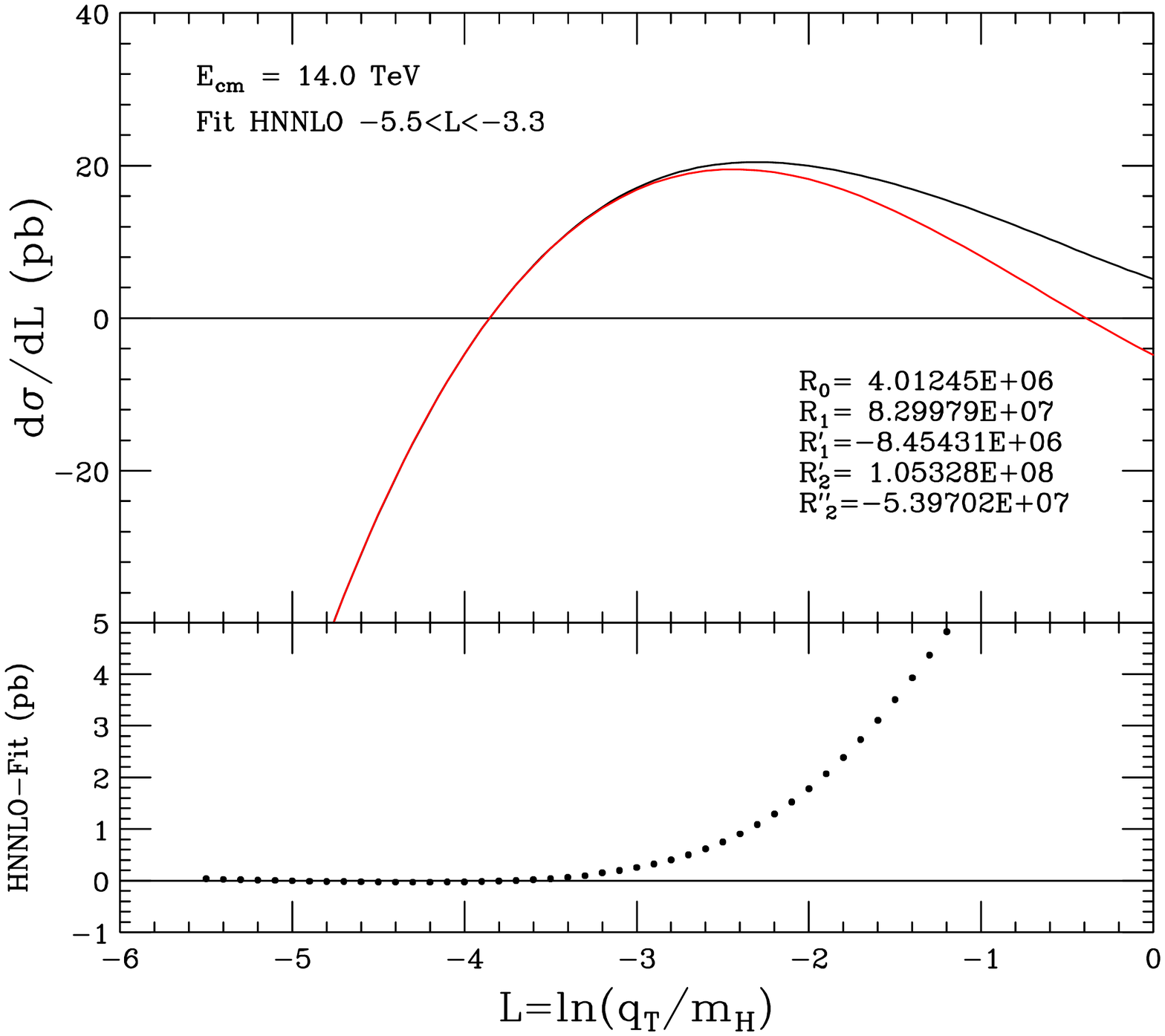,width=75mm}
\end{center}
\caption{Upper panels: fits to the logarithmic terms of the transverse-momentum
  ($q_T$) distribution in
  Higgs boson production at the LHC at 8 and 14 TeV.
Black: NLO data from HNNLO. Red: fit to data at $-5.5<L<-3.3$.
Lower panels: difference between the NLO data and the fits.}
\label{fig:fitPThigg126LHC}
\end{figure}

We evaluate the coefficients $R_0,R_1,R'_1$ explicitly from
Eqs.~(\ref{eq:RgNLO}), and then obtain the higher-order coefficients
$R'_2,R''_2$ from a fit to the Higgs transverse-momentum distribution, as
explained in Appendix~\ref{app:ptcomp}. These coefficients depend only 
on the parton distribution functions and the NLO coefficient functions
$C^{(1)}_{ga}$, which are the same for the $E_T$ and $q_T$
spectra. Using the MSTW 2008 NLO parton
distributions~\cite{Martin:2009iq}, we find the values indicated in
Fig.~\ref{fig:fitPThigg126LHC}, where the resulting fits are also
shown.

\begin{figure}
\begin{center}
\epsfig{file=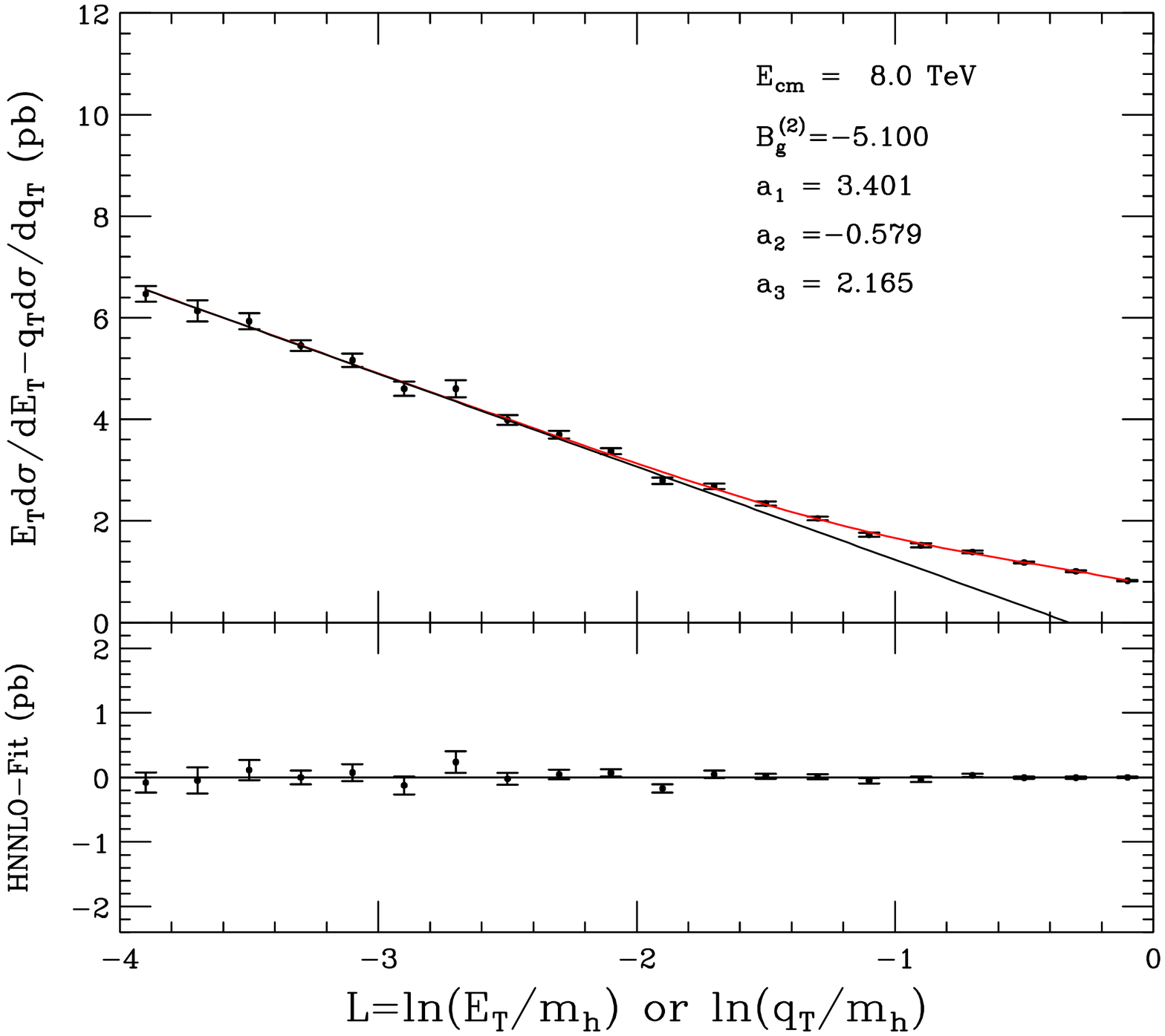,width=75mm}
\epsfig{file=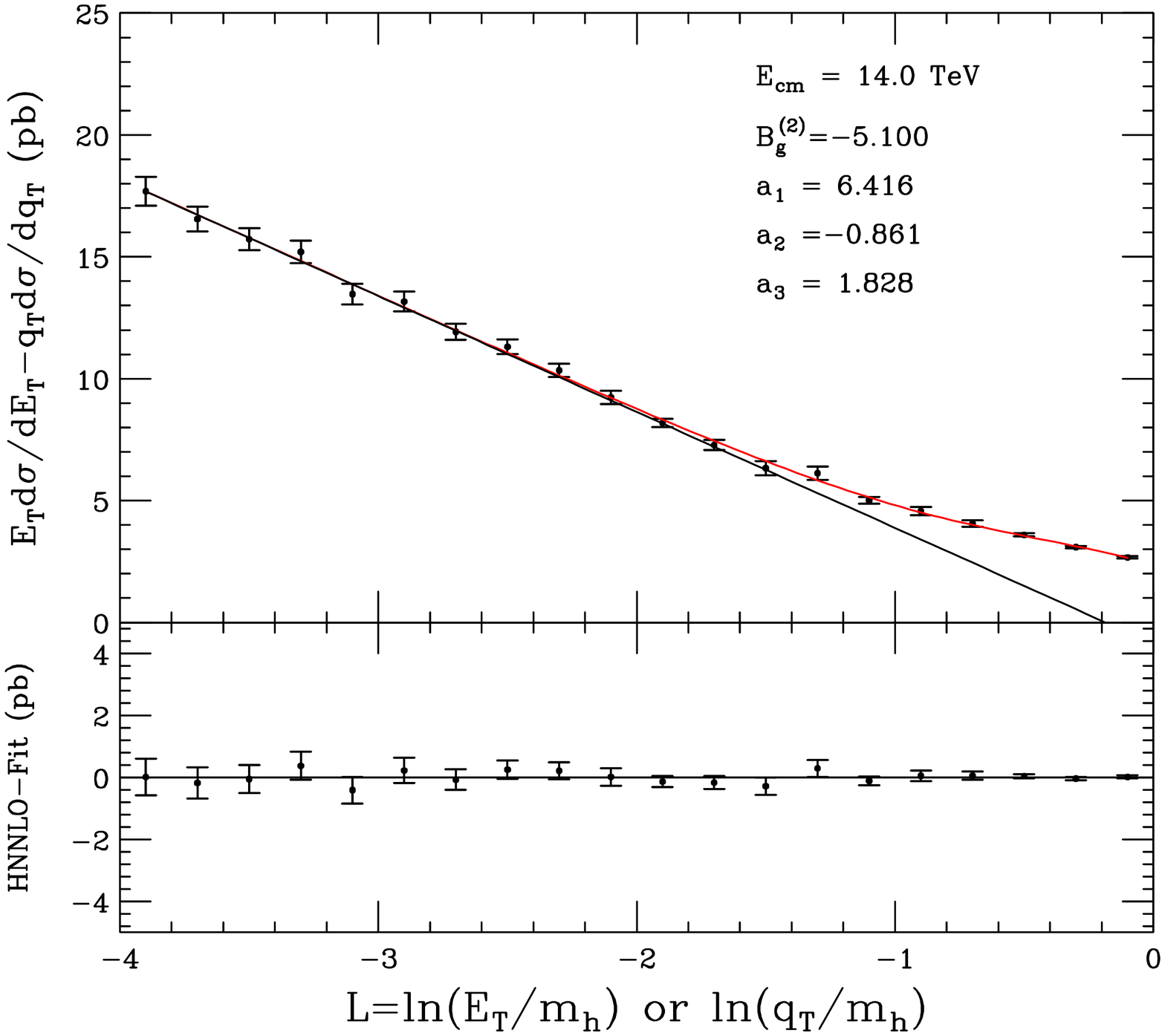,width=75mm}
\end{center}
\caption{Upper panels: fits to the difference between the transverse-energy ($E_T$) and
  transverse-momentum  ($q_T$) distribution in
  Higgs boson production at the LHC at 8 and 14 TeV.
Points: NLO data from HNNLO. Red: fit to data. Black: logarithmic
terms only.  Lower panels: difference between the NLO data and the fits.} 
\label{fig:fitETPThigg126LHC}
\end{figure}

\subsection{Matching to NLO}\label{sec:matchNLO}
The NLO prediction for the transverse-energy distribution is conveniently
obtained from the known NLO transverse-momentum distribution by adding the
difference between the two distributions, obtained
from Higgs plus two-jet production at leading order.  Given the value of
$B_g^{(2)}$ for the transverse-energy distribution, the NLO prediction
(\ref{eq:EtPtdif}) for the difference at small $E_T$ is independent of the
fitted parameters $R'_2,R''_2$.  From HNNLO~\cite{Catani:2007vq,Grazzini:2008tf} data on this
quantity at 8 and 14 TeV, shown in Fig.~\ref{fig:fitETPThigg126LHC}, we find
consistent best-fit $B_g^{(2)}$ values of $-4.5\pm 2.1$ and $-6.0\pm 2.6$,
respectively, with weighted average $B_g^{(2)}=-5.1\pm 1.6$.  This is
significantly different from the value of $\bB_g^{(2)}=26.8$ given by
Eqs.~(\ref{eq:Agetc}) for the transverse-momentum distribution.  We will use
$B_g^{(2)}=-5.1$ from now on.

 Away from the small-$E_T$ region, the NLO data are then well described by a
 parametrization of the form \beq\label{eq:EtPtfit}
 \left[\frac{d\sigma_H}{dE_T}-
   \left.\frac{d\sigma_H}{dq_T}\right|_{q_T=E_T}\right]_{\rm NLO}
 =\mbox{Logarithmic terms} + \frac{a_1 E_T}{m_H(m_H+a_2 E_T)+a_3 E_T^2}\,, \eeq
 as shown by the red curves in Fig.~\ref{fig:fitETPThigg126LHC}, with the
 parameter values shown.

To match the resummed and NLO $E_T$ distributions, we have to subtract
the NLO logarithmic terms (\ref{eq:ETNLO}), which are already included in the
resummation, and replace them by the full NLO result:
\beq\label{eq:dsigmas}
\frac{d\sigma_H}{dE_T} =\left[\frac{d\sigma_H}{dE_T}\right]_{\rm resum}
-\left[\frac{d\sigma_H}{dE_T}\right]_{\rm resum,NLO}
+\left[\frac{d\sigma_H}{dE_T}\right]_{\rm NLO}\,.
\eeq

\subsection{Results}\label{sec:res}
In the following we present numerical results for our resummed
calculation of the $E_T$ distribution at the LHC.
Our resummed results are obtained by using Eq.~(\ref{eq:dsigmas}): the resummed
component is evaluated by including the coefficients $C_{ga}^{(1)}$ in
Eq.~(\ref{eq:Agetc}), and the functions $g_1$, $g_2$ and $g_3$ of
Eqs.~(\ref{eq:gs}). The required coefficients $A_g^{(1)}$, $A_g^{(2)}$ and
$B_g^{(1)}$ are given in Eq.~(\ref{eq:Agetc}). For the coefficient $B_g^{(2)}$ we
use the numerical value extracted in Sec.~\ref{sec:matchNLO}.  The
unknown coefficient $A_g^{(3)}$ is neglected. We will comment later on
the numerical impact of the missing $A_g^{(3)}$ and $C_{ga}^{(2)}$ coefficients.

\begin{figure}
\begin{center}
\epsfig{file=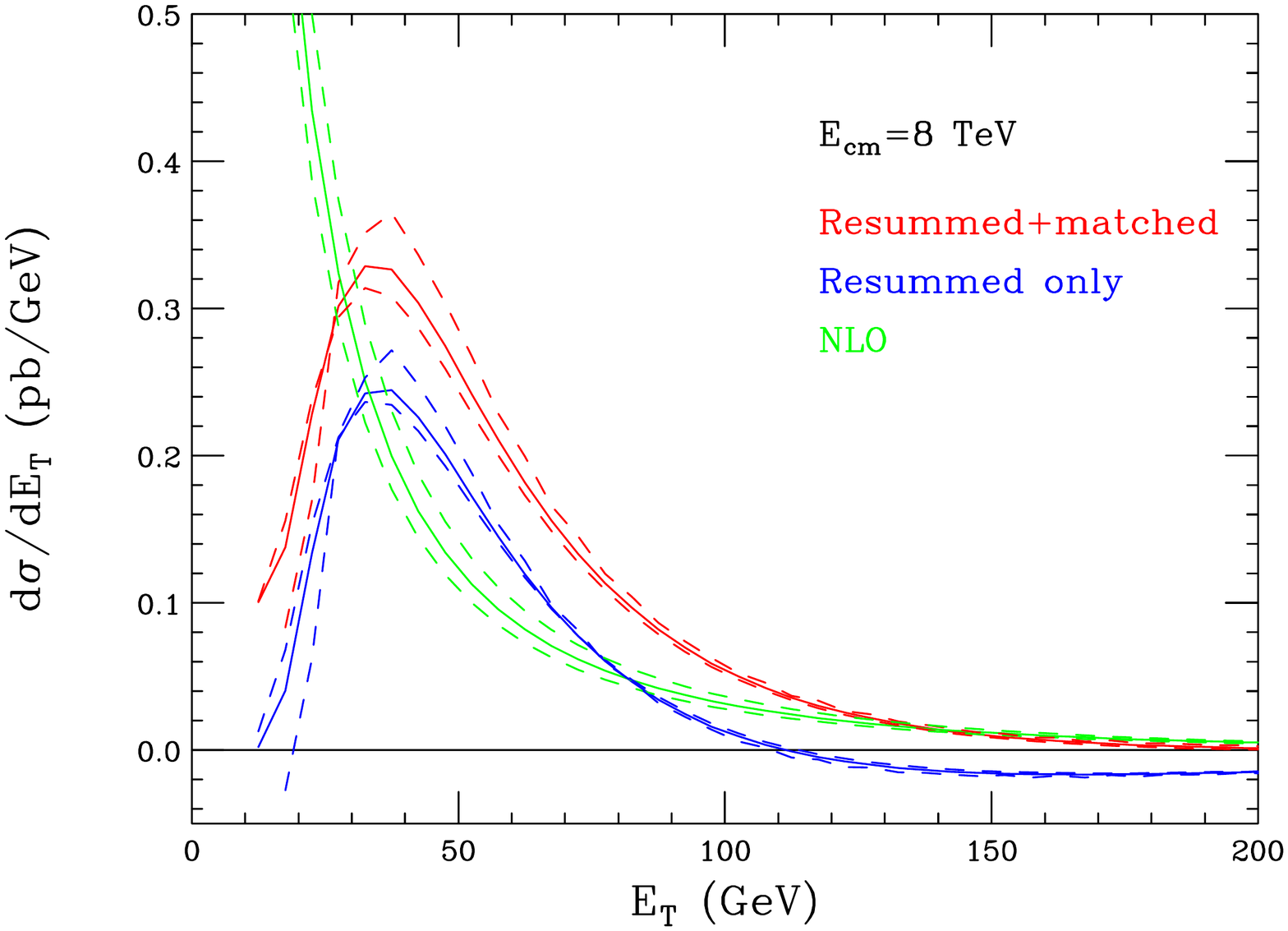,width=75mm}
\epsfig{file=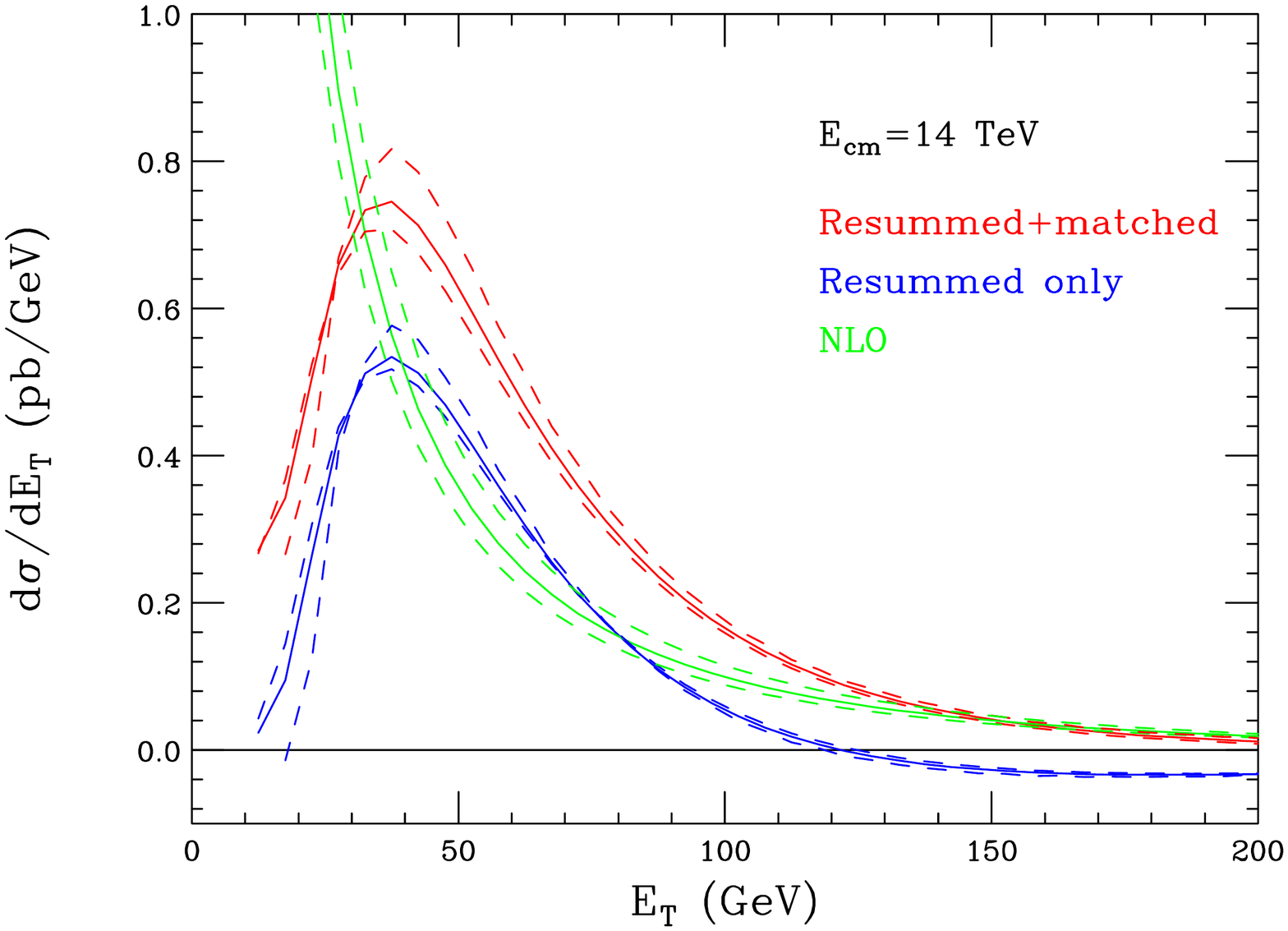,width=75mm}
\end{center}
\caption{Transverse-energy distribution in
  Higgs boson production at the LHC at 8 and 14 TeV.
Blue: resummed only.
Red: resummed and matched to NLO.
Green: NLO only.  The solid curves correspond to renormalization scale
$m_H$,  the dashed to $2 m_H$ and $m_H/2$.}
\label{fig:sumEThigg126LHC}
\end{figure}
\begin{figure}
\begin{center}
\epsfig{file=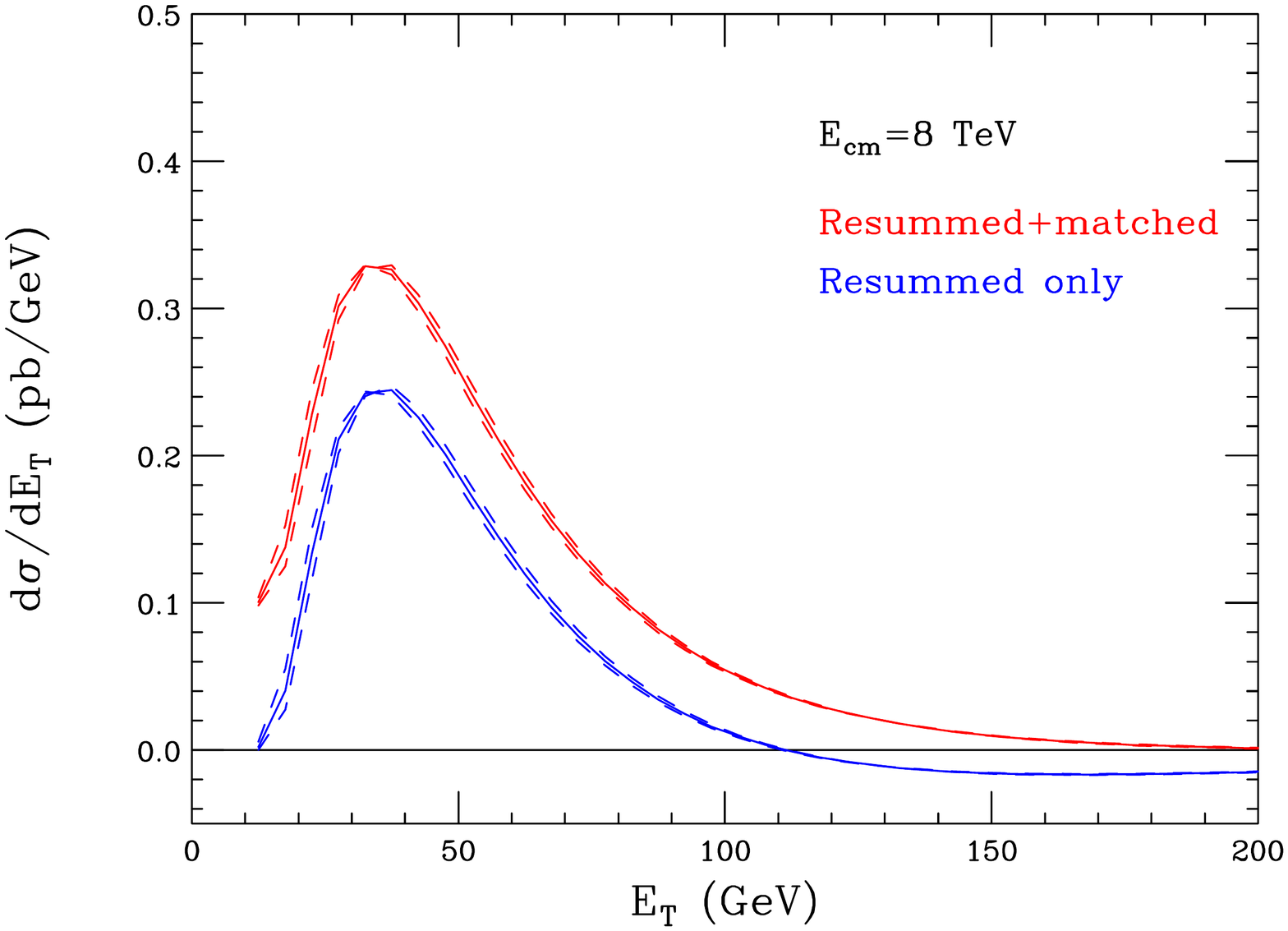,width=75mm}
\epsfig{file=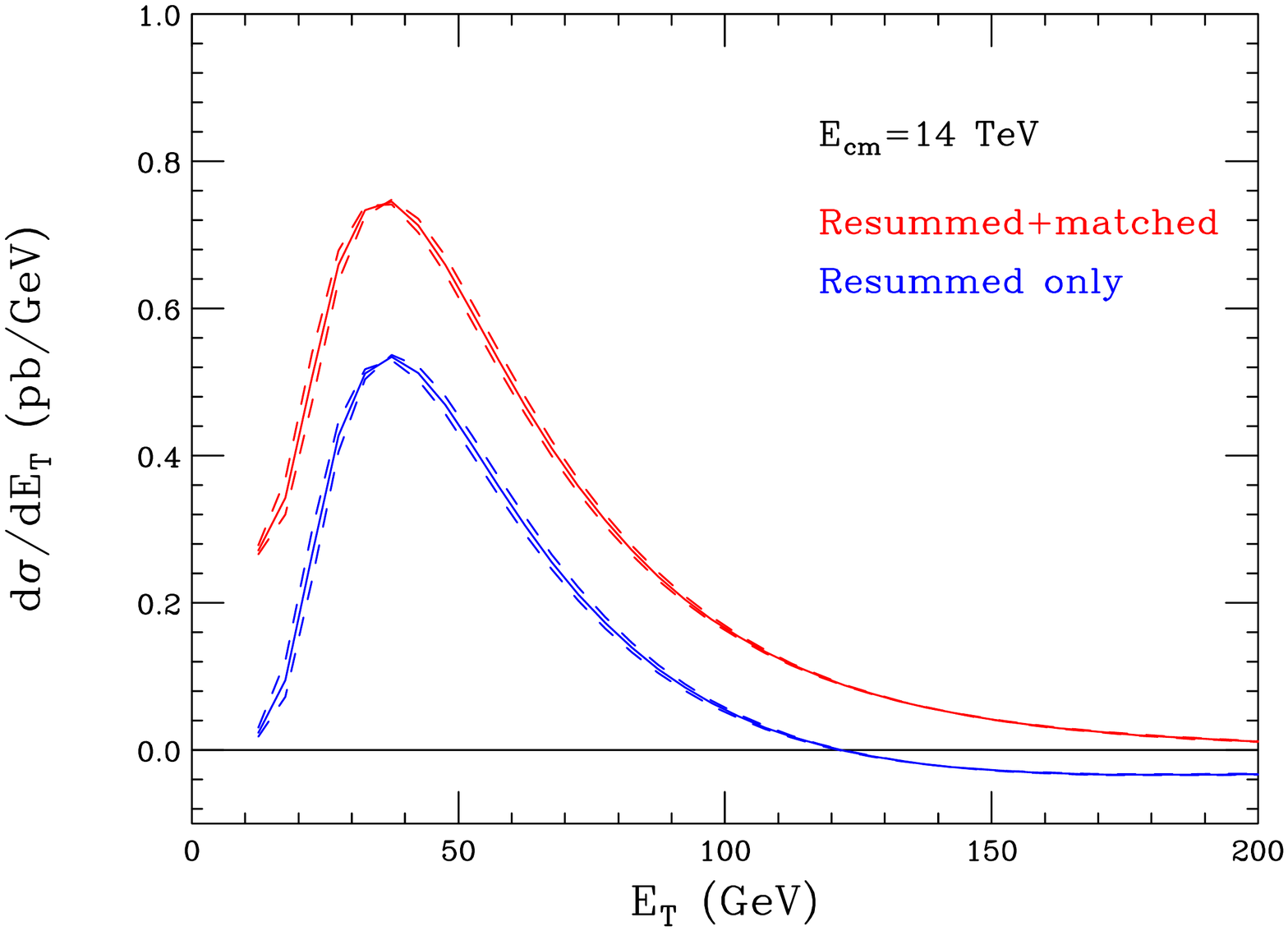,width=75mm}
\end{center}
\caption{Transverse-energy distribution in
  Higgs boson production at the LHC at 8 and 14 TeV.
Blue: resummed only.
Red: resummed and matched to NLO.
The solid curves correspond to  $A_g^{(3)}=0$,
the dashed to $A_g^{(3)}=\pm 30$.}
\label{fig:sumEThigg126LHCA3}
\end{figure}
\begin{figure}
\begin{center}
\epsfig{file=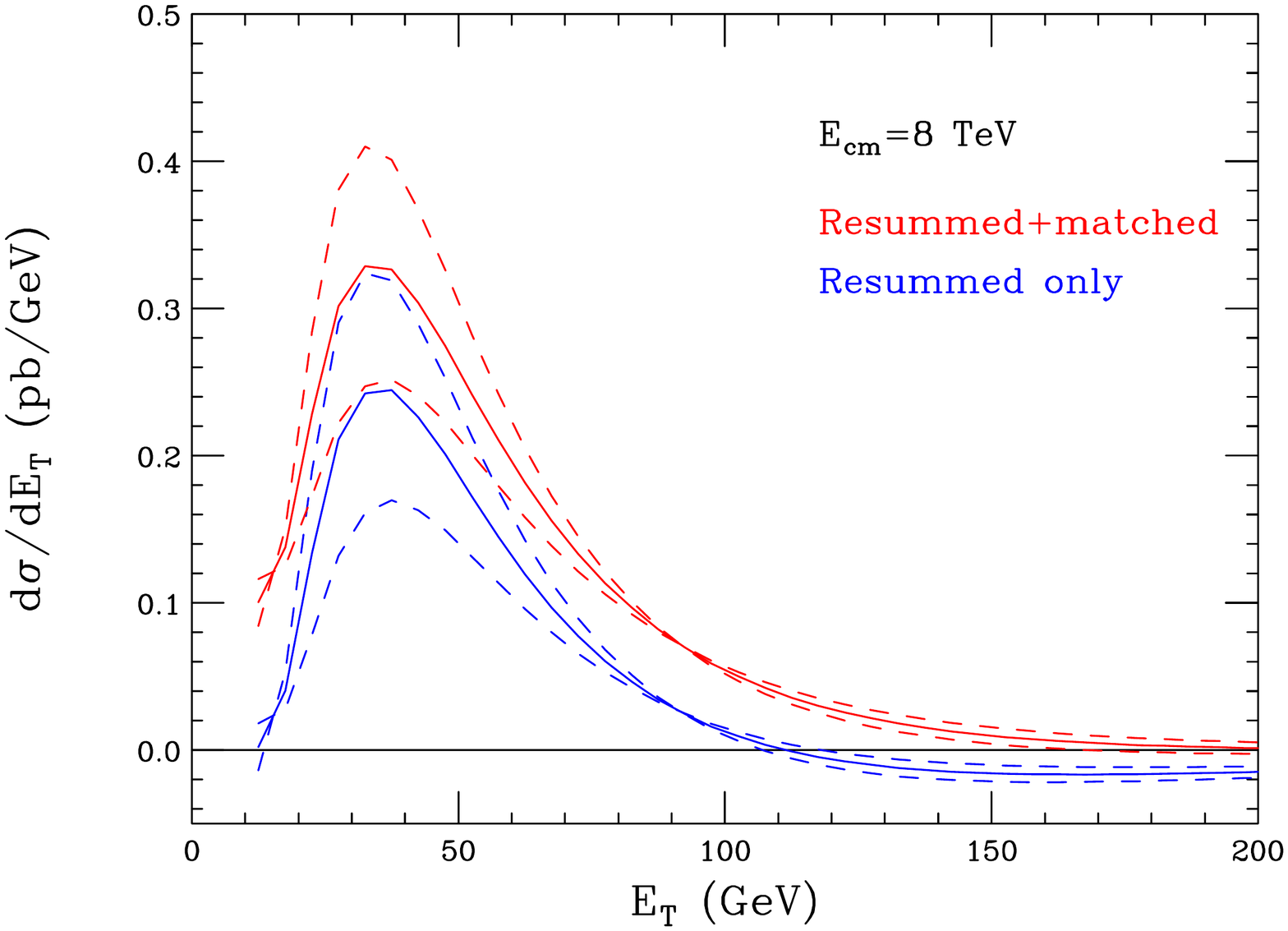,width=75mm}
\epsfig{file=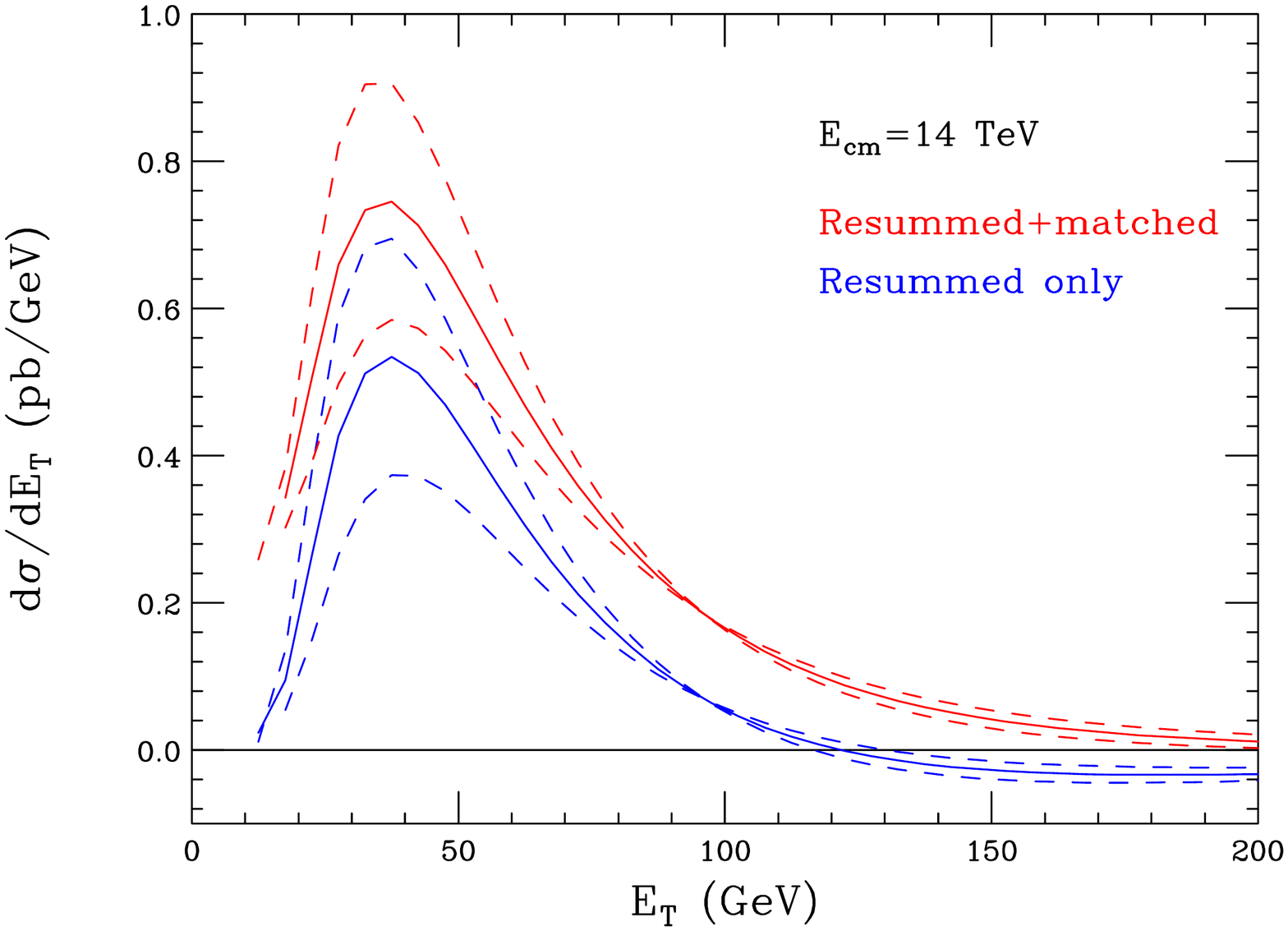,width=75mm}
\end{center}
\caption{Transverse-energy distribution in
  Higgs boson production at the LHC at 8 and 14 TeV.
Blue: resummed only.
Red: resummed and matched to NLO.
The solid curves correspond to  $C_g^{(2)}=0$,
the dashed to $C_g^{(2)}=\pm 115.5$ (see text).}
\label{fig:sumEThigg126LHCC2}
\end{figure}
The resulting resummed and matched $E_T$ distributions at the LHC at 8 and 14
TeV are shown in Fig.~\ref{fig:sumEThigg126LHC}. For all these predictions we
use the best-fit value $B_g^{(2)}=-5.1$ found from the NLO data. The
distribution peaks at around $E_T=35$ GeV at both centre-of-mass
energies, considerably above the peak in the Higgs transverse-momentum
distribution, around $q_T=12$ GeV~\cite{Bozzi:2005wk}.
Thus in the peak region of $E_T$ the resummed logarithms
are not so dominant as in the corresponding region of $q_T$.
On the other hand, the fixed-order NLO prediction is rising rapidly 
and unphysically towards smaller values of $E_T$.\footnote{At very
  small $E_T$ it turns over and tends to $-\infty$, as seen in
  Fig.~\ref{fig:fitPThigg126LHC}.}

The purely
resummed distribution becomes slightly negative at small and large $E_T$, which
is also unphysical.  The effect of matching is to raise the distribution to positive
values, close to the fixed-order prediction at high $E_T$.  The
matched prediction is still somewhat unstable at small $E_T$, owing to
the delicate cancellation of diverging logarithmic terms.  The
behaviour at large $E_T$ has been significantly improved compared to the results
of Ref.~\cite{Papaefstathiou:2010bw} where matching was only performed to
leading order.  The renormalization scale dependence is comparable to that of
the NLO result above and around the peak of the distribution, but changes sign
at lower $E_T$, with smaller scales giving a lower cross section there
and a peak that is higher and shifted to slightly larger $E_T$.

As a result of the unitarity condition (\ref{eq:unit}) and the
matching to fixed order, the cross section integrated over all $E_T$
should be equal to the NNLO inclusive Higgs cross section, within the
uncertainties due to the unknown coefficients. Integrating up to
$E_T=760$ GeV, we obtain resummed matched cross sections of 18.4 pb
and 47.8 pb at 8 and 14 TeV, respectively, which compare well with the
NNLO inclusive values of 18.22 pb and 47.28 pb, computed with the
same NLO PDFs and two-loop $\as$.

As mentioned above, the leading terms that are neglected
in our analysis correspond to the coefficient $A_g^{(3)}$ in
Eq.~(\ref{aexp}) and the coefficient functions $C_{ga}^{(2)}(z)$ in (\ref{cexp}).
In Fig.~\ref{fig:sumEThigg126LHCA3} we show the sensitivity
of the prediction to $A_g^{(3)}$, assuming a value similar in
magnitude to that used for the $q_T$ distribution in
Ref.~\cite{Bozzi:2005wk}.  As was found there, the effect of
including this coefficient is small.

The NNLO coefficient functions $C^{(2)}_{ga}(z)$
were computed for the Higgs transverse-momentum distribution
in Ref.~\cite{Catani:2011kr}, where it was found that their dominant effect
could be approximated as
\beq
 {\overline C}^{(2)}_{ga}(z) \approx  {\overline C}^{(2)}_g
 \delta_{ag}\,\delta(1-z)
\eeq
where ${\overline C}^{(2)}_g=115.5$.  In Fig.~\ref{fig:sumEThigg126LHCC2}
we show the effects of assuming the same form and magnitude for the
corresponding coefficient in $E_T$ resummation.   We see that the
effect is larger than that of $A_g^{(3)}$ and of renormalization scale
variation.  Thus in this case uncertainties in the higher-order
coefficient functions provide a more conservative error estimate
that the usual range of scale variation.

\section{Monte Carlo studies}\label{sec:MC}
Up to this point we have only considered the perturbative contributions to the
$E_T$ distribution that arise due to initial state radiation (ISR).  However,
there are important contributions to the $E_T$ originating from non-perturbative
effects such as hadronization and the underlying event (UE). Moreover, the
distributions can be altered further because of cuts imposed either due to the
detector geometry, or to accommodate the experimental analyses.

All of the aforementioned effects on the $E_T$ distribution are challenging to
predict analytically. Therefore, we make the assumption that the
kinematics of the process, apart from the UE, is governed predominantly by the shape of
the $E_T$ distribution. Under this assumption, one can reweight the
parton-level $E_T$ of a Monte Carlo event generator (i.e. with the UE
and hadronization turned off), to the one
calculated analytically, and use the phenomenogical models of the
Monte Carlo to estimate the features of the effects. 

Taking into account the non-perturbative and detector geometry effects, one may
construct a quantity that is, at least in principle, close to what can
be measured experimentally:
\begin{equation}\label{eq:expdef}
E_T = \sum_{\substack{|\eta_i| < \eta^c \\
        |\mathbf{p}_{Ti}| > p_{T}^c  }}
   \hspace{-2mm}|\mathbf{p}_{Ti}| \;,
\end{equation}
where the sum is taken over the hadrons in the event, with $p_{T}^c$
and $\eta^c$ their minimum transverse momentum and maximum pseudorapidity
respectively. The effect of these cuts will be investigated below.\footnote{The $E_T$ distribution may also be constructed using
jets. It is not feasible, however, to reconstruct the parton- or
hadron-level distributions from the jet-level distributions. }

\subsection{$E_T$ at parton level}
\begin{figure}[!htb]
    \includegraphics[width=0.38\linewidth,
    angle=90]{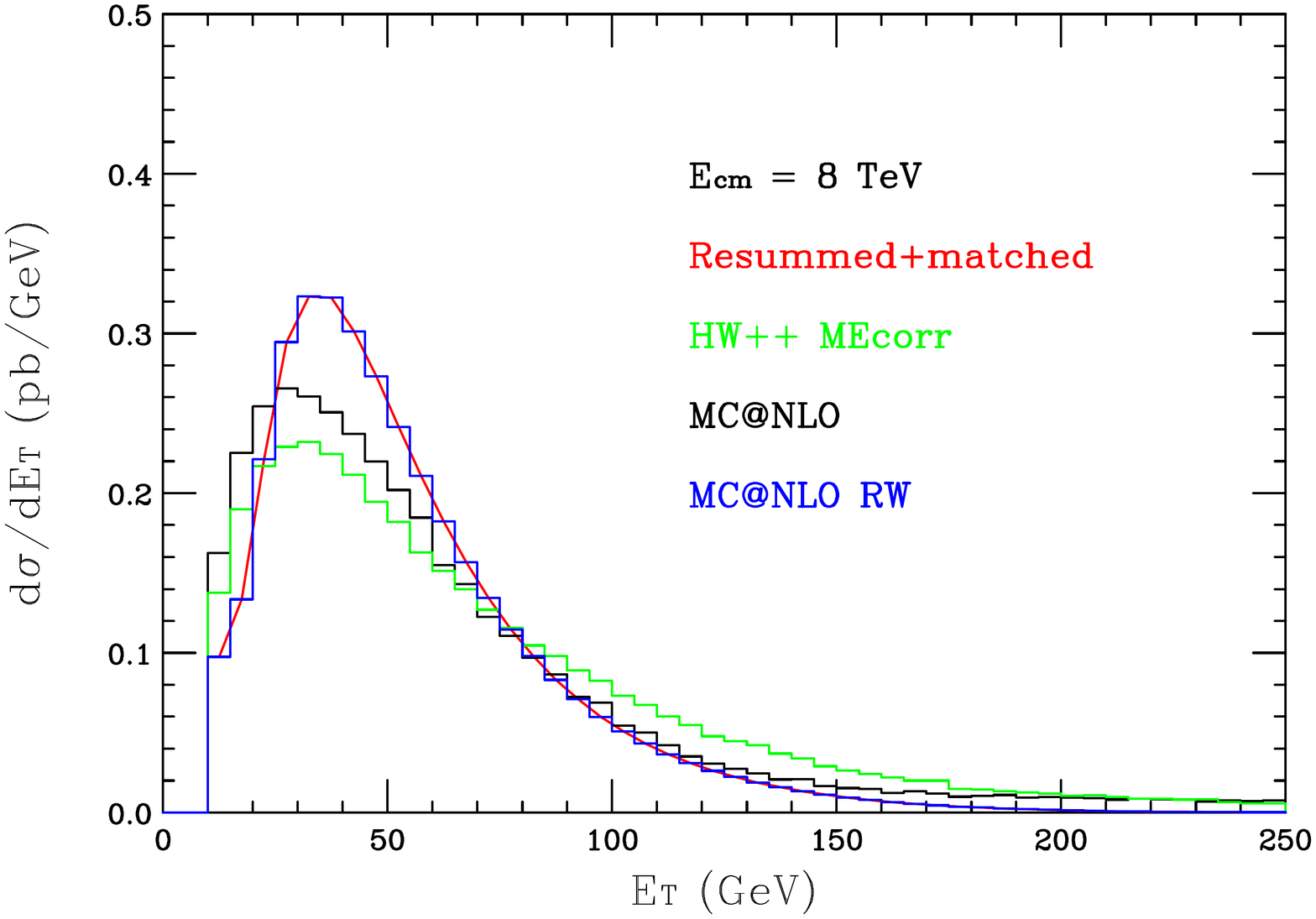}
    \includegraphics[width=0.38\linewidth, angle=90]{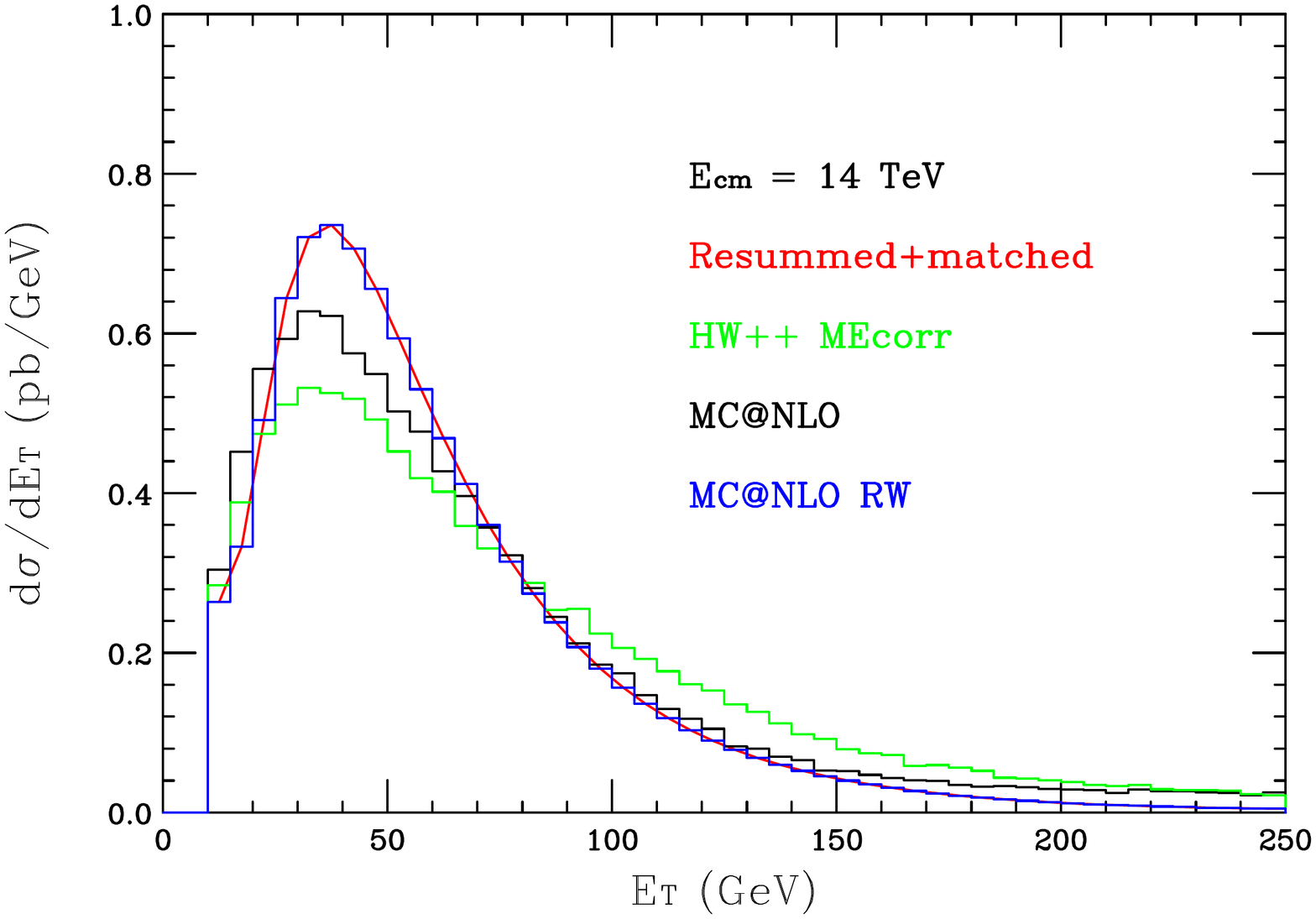}
  \caption{Parton-level transverse-energy distribution in
  Higgs boson production at the LHC at 8 and 14 TeV.
Red: resummed and matched to NLO.
Green: \HW\ with Matrix Element correction, no
reweighting. 
Black: \aMC+\HW\ before reweighting.
Blue: \aMC+\HW\ after reweighting.}
  \label{fig:mcnlo}
\end{figure} 

Here we employ the \HW\ general-purpose event generator (version
2.6.3)~\cite{Bahr:2008pv, Arnold:2012fq} in conjunction with events generated
using \aMC~\cite{Frixione:2002ik,Frederix:2011ss}. For purposes of comparison with alternative descriptions of the parton shower, hadronization and the underlying event, we additionally verify the \HW\ results using the \PY\ event generator~\cite{Sjostrand:2006za,Sjostrand:2007gs}. The distributions found using \PY\ are very similar to those obtained with \HW\ and thus we defer them to Appendix~\ref{app:pythia}.

\begin{figure}[!htb]
    \hspace{0.3cm}
    \includegraphics[width=0.40\linewidth]{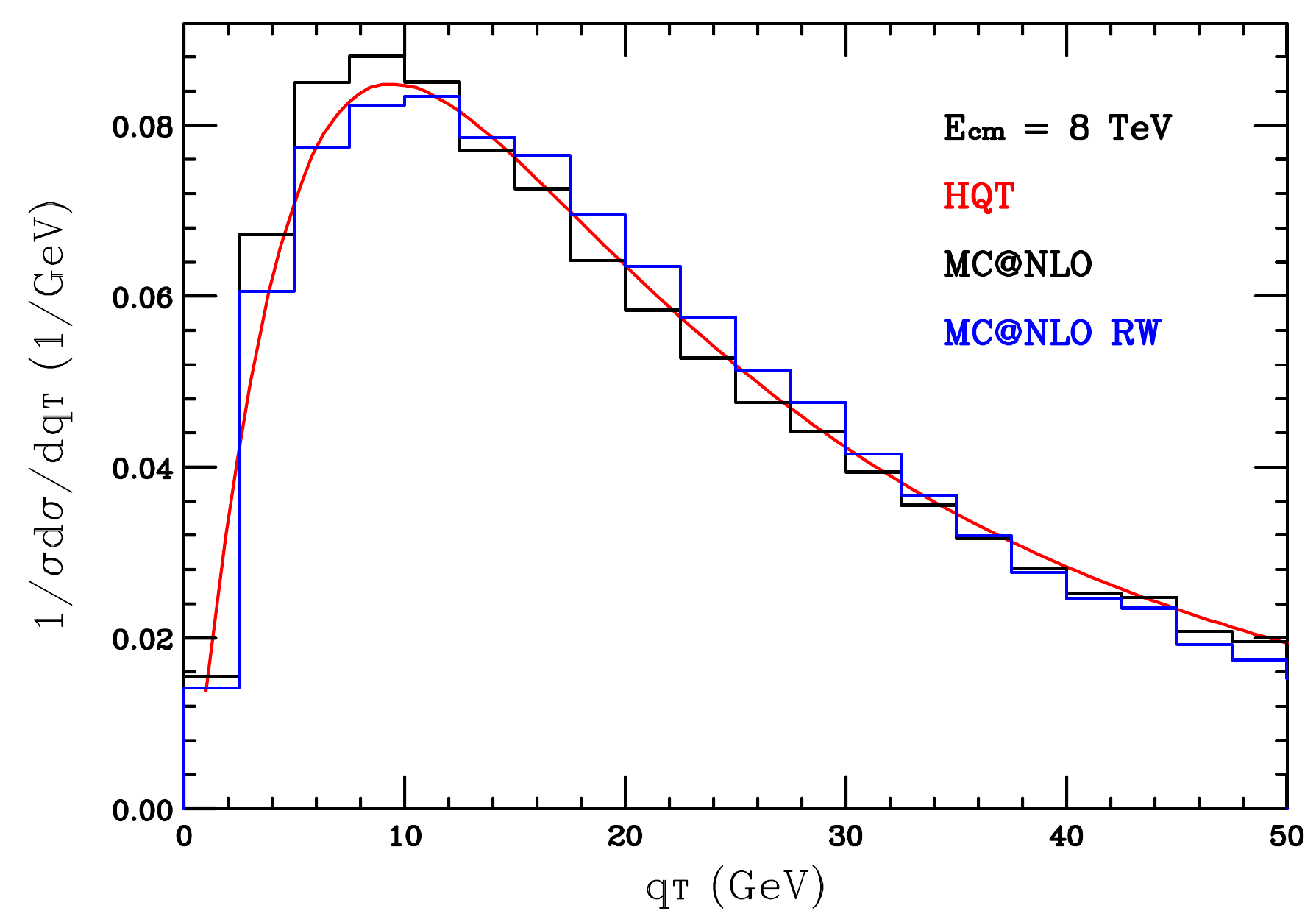}
    \hspace{1.4cm}
    \includegraphics[width=0.40\linewidth]{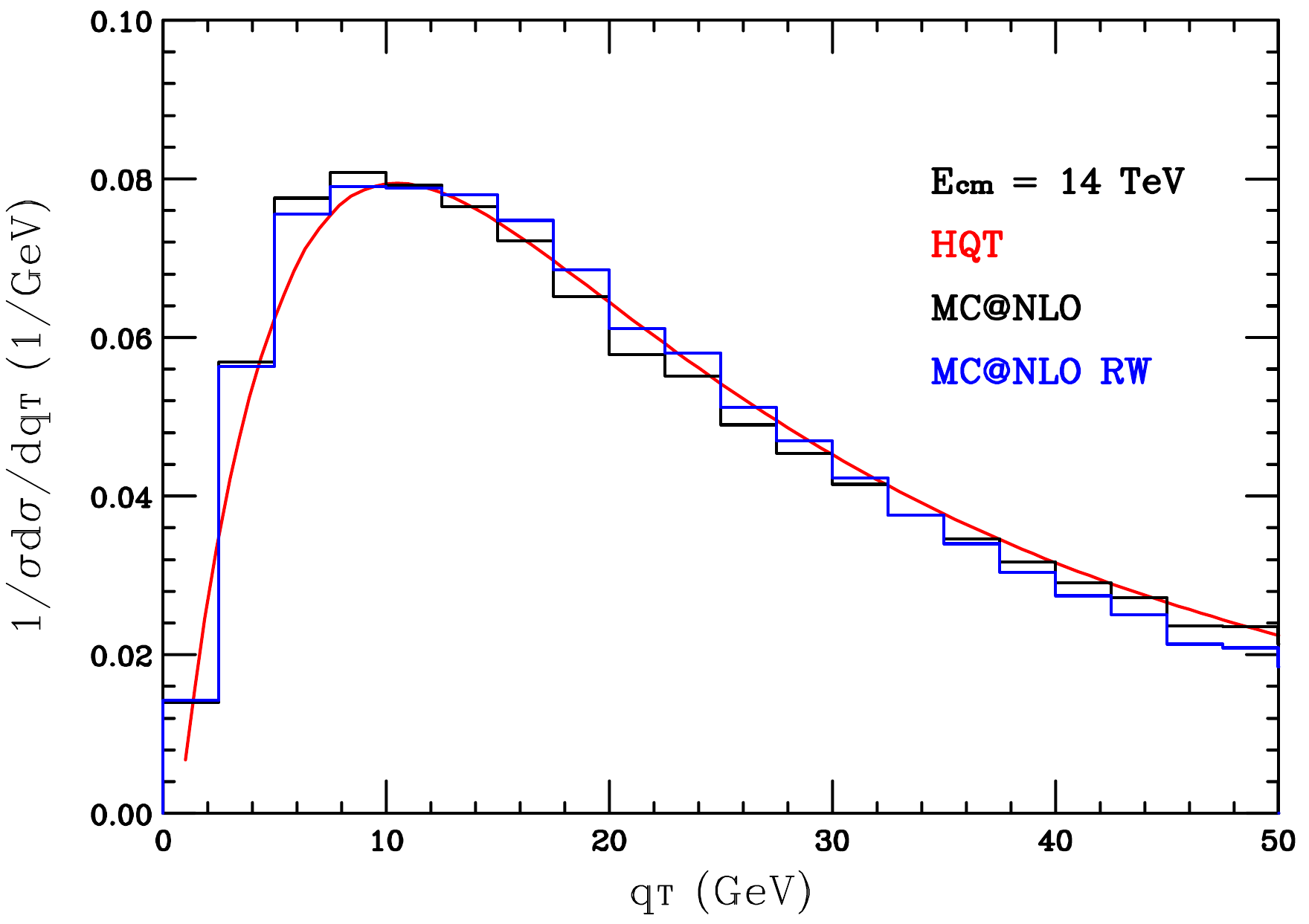}
  \caption{Higgs boson transverse-momentum distribution at the LHC at 8 and 14 TeV.
Red: \texttt{HQT} calculation.
Black: \aMC+\HW\ before reweighting.
Blue: \aMC+\HW\ after reweighting.}
  \label{fig:qtatpl}
\end{figure}

\begin{figure}[!htb]
    \includegraphics[width=0.38\linewidth,
    angle=90]{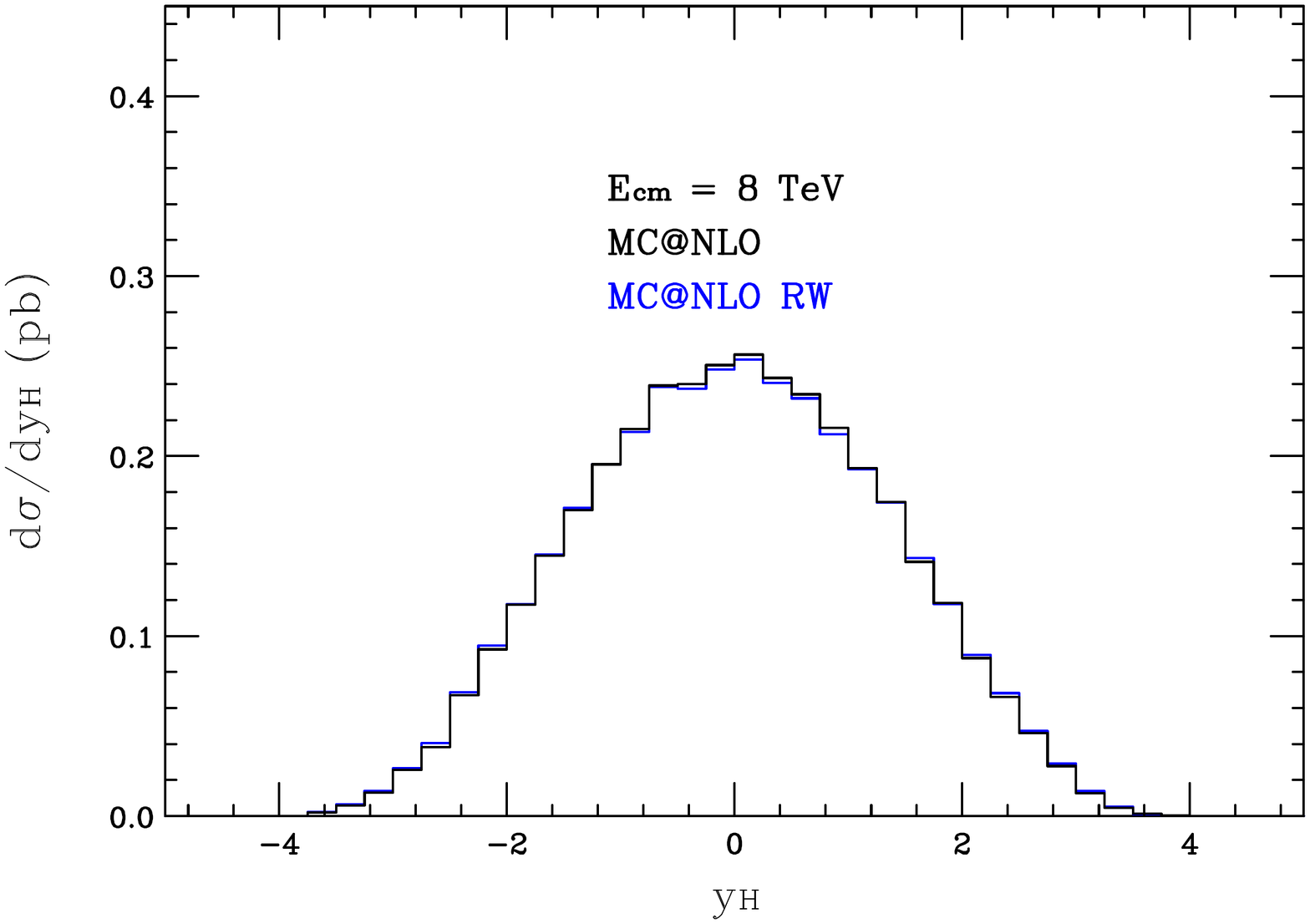}
    \includegraphics[width=0.38\linewidth, angle=90]{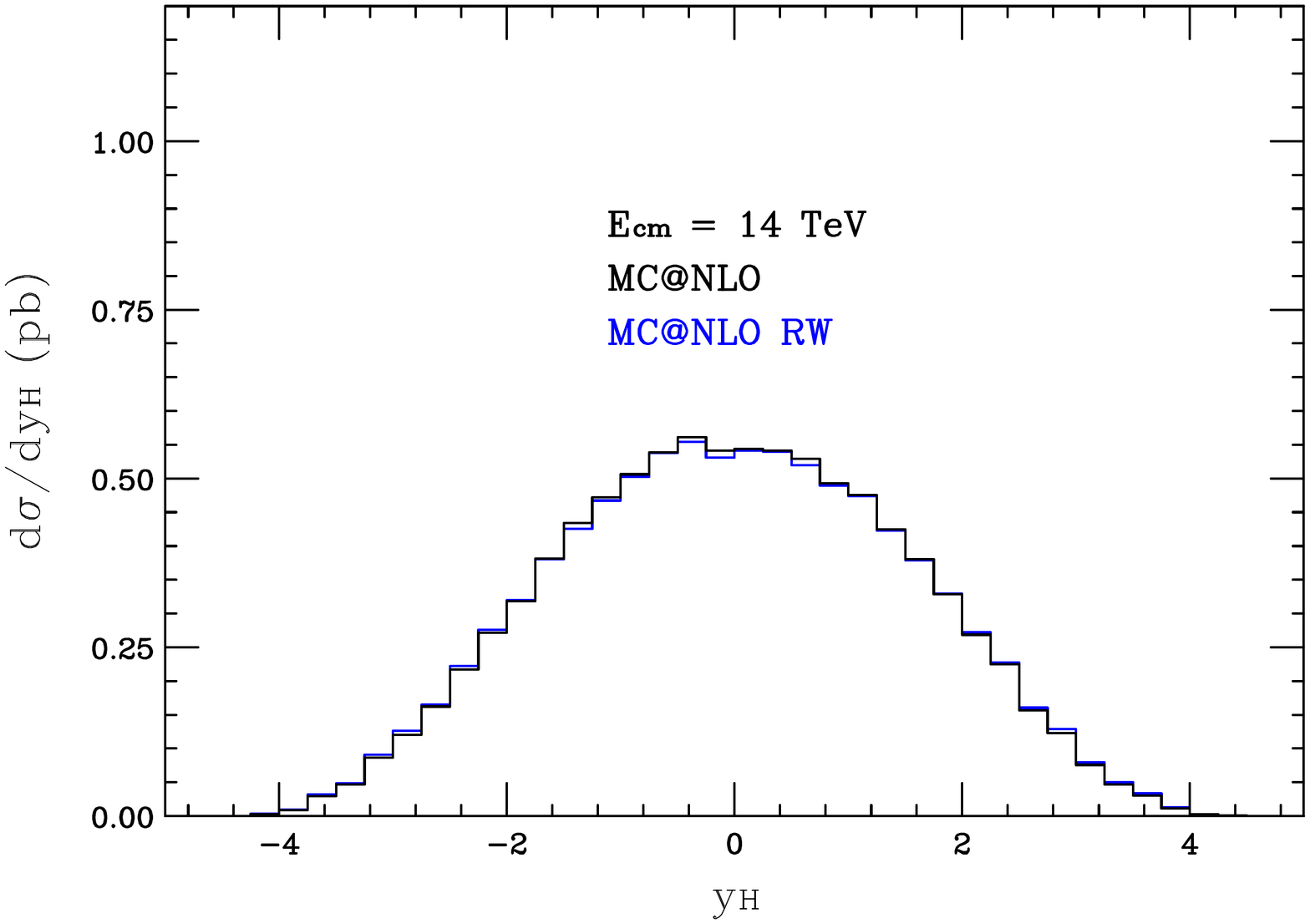}
  \caption{Higgs boson rapidity distribution at the LHC at 8 and 14 TeV.
Black: \aMC+\HW\ before reweighting.
Blue: \aMC+\HW\ after reweighting.}
  \label{fig:etaatpl}
\end{figure}

Use of \aMC\ ensures
correct treatment of the NLO inclusive cross section matched to parton
showers without double counting.  We assign a new weight to each
event so as to reproduce the analytic resummed and matched distributions shown
in Fig.~\ref{fig:sumEThigg126LHC}. For completeness, we show in
Fig.~\ref{fig:mcnlo} the resulting $E_T$ distributions before and after the
reweighting, at parton level, demonstrating that this procedure
reproduces the analytic result. Higgs boson production using the
internal \HW\ implementation,
which includes Matrix Element (ME) corrections\footnote{The ME corrections
  include the higher-order tree-level contributions $gg \rightarrow hg$, $qg
  \rightarrow hq$, $g\bar{q} \rightarrow h\bar{q}$ and $q\bar{q} \rightarrow
  hg$.}, is also shown on the figure. The ME-corrected $E_T$ distribution has a
lower peak and consequently falls off more slowly at higher $E_T$ than the
MC@NLO case. In Fig.~\ref{fig:qtatpl} we show the Higgs boson transverse-momentum distribution, $q_T$, before and after applying the reweighting
procedure, compared to the equivalent distribution obtained by the \texttt{HQT}
program~\cite{Bozzi:2005wk, deFlorian:2011xf}. Evidently, the MC@NLO
distribution agrees already quite well with the \texttt{HQT} prediction
before reweighting. The reweighting procedure improves agreement in
the peak region but makes the distribution fall off faster at high
$q_T$, which is consistent with the change in shape observed in
Fig.~\ref{fig:mcnlo}. Figure~\ref{fig:etaatpl} shows the rapidity distribution
of the Higgs boson, $y_H$, before and after the reweighting, clearly showing
that the effect on this distribution is negligible, thus verifying that the
reweighting does not alter physics in the forward direction.

\subsection{$E_T$ at hadron level}
\begin{figure}[!htb]
    \includegraphics[width=0.38\linewidth,
    angle=90]{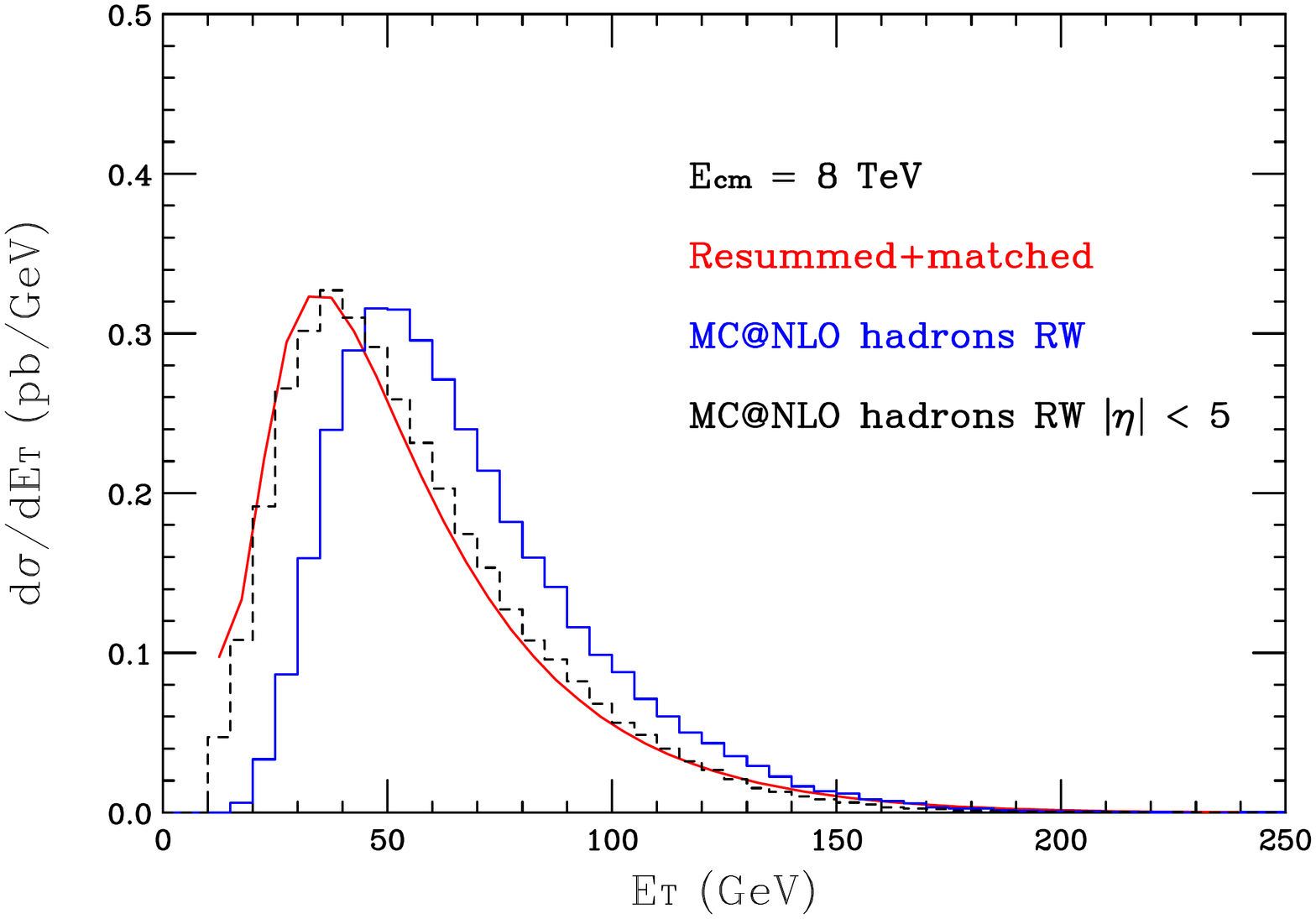}
    \includegraphics[width=0.38\linewidth, angle=90]{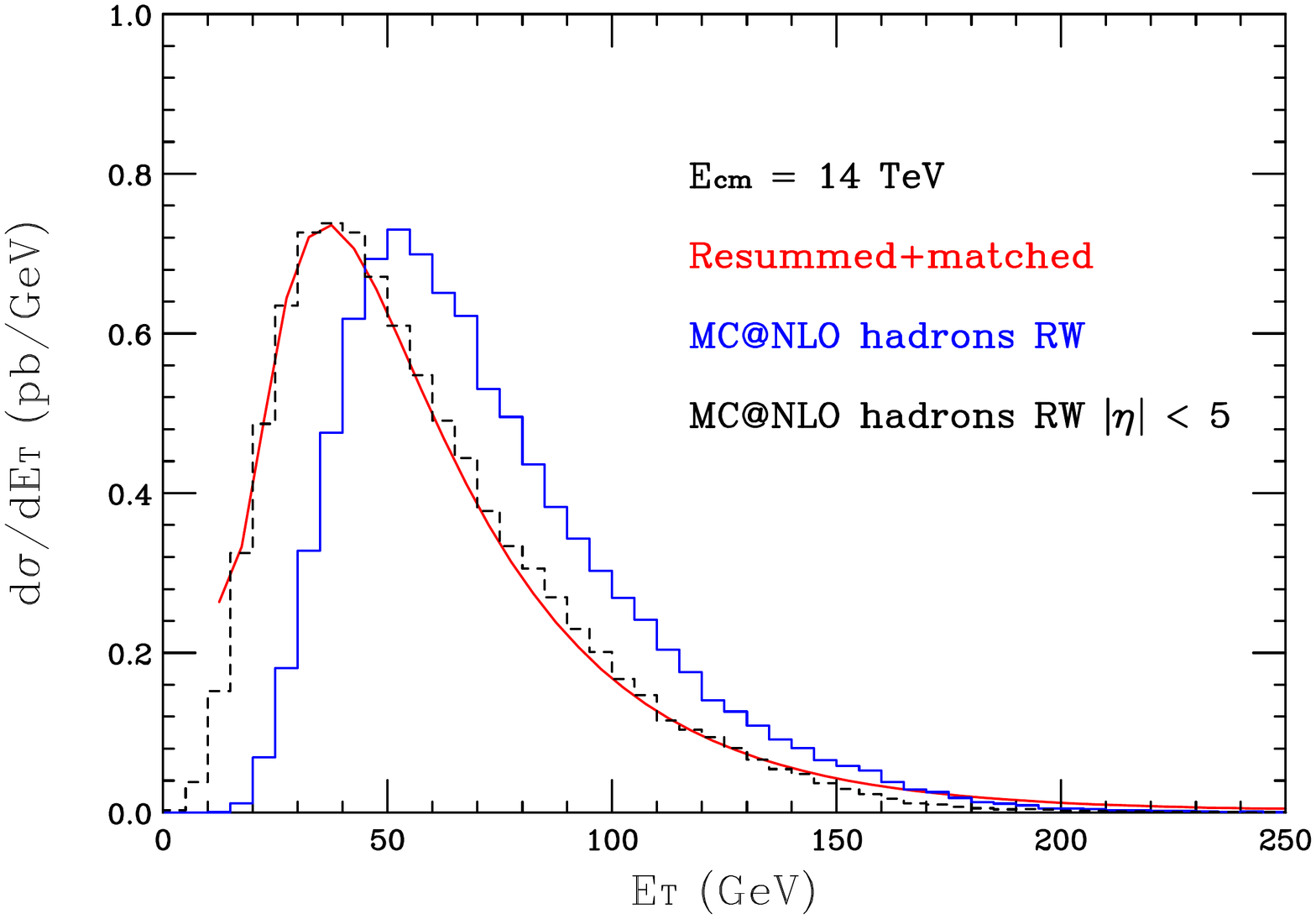}
  \caption{Hadron-level transverse-energy distribution in
  Higgs boson production at the LHC at 8 and 14 TeV.
Red: resummed and matched to NLO.
Blue: \aMC+\HW\ after reweighting.
Black dashes: \aMC+\HW\ after reweighting, particles
restricted to lie within pseudorapidity $|\eta| < 5$.
}
  \label{fig:mcnlo_hadron}
\end{figure}

The effects of hadronization can be studied by enabling the cluster
hadronization model in the \HW\ event generator in
conjunction with the reweighting. The default parameters of the
hadronization model available in \HW\  version 2.6.3 were used. The
effect is to shift the peak of the distribution to higher $E_T$, by
about 15~GeV at both 8~TeV and 14~TeV, as shown in
Fig.~\ref{fig:mcnlo_hadron}. The effect of hadronization on the $E_T$
distribution can be compensated almost completely in this range of values by imposing a pseudorapidity cut on the hadrons
contributing to the $E_T$, allowing only hadrons within $|\eta| < 5$ to
enter. The resulting distributions after this cut are also shown in
Fig.~\ref{fig:mcnlo_hadron}. We note that including the restriction on
hadrons of $|\eta| < 5$ approximately corresponds to the experimental
detector coverage of the ATLAS and CMS detectors.

\subsection{Inclusion of the underlying event}
The underlying event (UE) is thought to arise due to secondary multiple
interactions between the colliding hadrons. The model present in
\HW\ is based on the eikonal model formulated in
Refs.~\cite{Durand:1988ax, Butterworth:1996zw, Borozan:2002fk}.  The
underlying event activity is treated as additional semi-hard and soft partonic
scatters.  In this version, a model of colour reconnection has been added to {\tt
  HERWIG++}, based on the idea of colour preconfinement, which provides an
improved description of underlying event data at the LHC~\cite{Gieseke:2012ft}.

The effect of the UE on the $E_T$ distributions is severe, making them
much broader and moving the peak to much higher values of $E_T$. This was investigated in Ref.~\cite{Papaefstathiou:2010bw} at parton level, where it was
shown that in the \HW\ model the $E_T$ distribution for
the partons originating from the UE is approximately independent of
the nature of the hard process.
This distribution was fitted with a Fermi distribution and was shown to
reproduce the total distribution after convolution with the
perturbative result.\footnote{This approach, however, predicts
  distributions only at parton-level.}

\begin{figure}[!htb]
 
    \includegraphics[width=0.38\linewidth,angle=90]{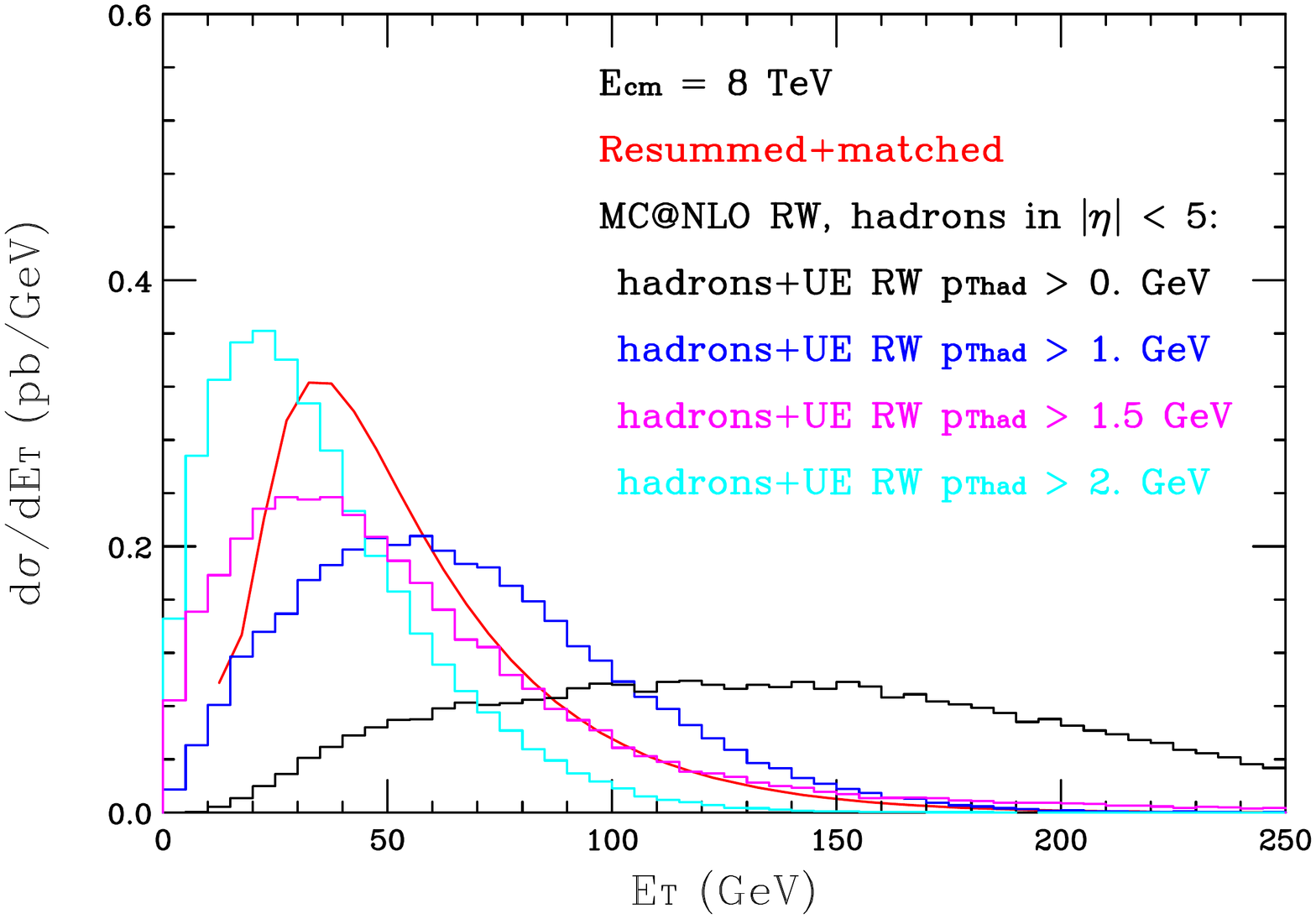}
    \includegraphics[width=0.38\linewidth, angle=90]{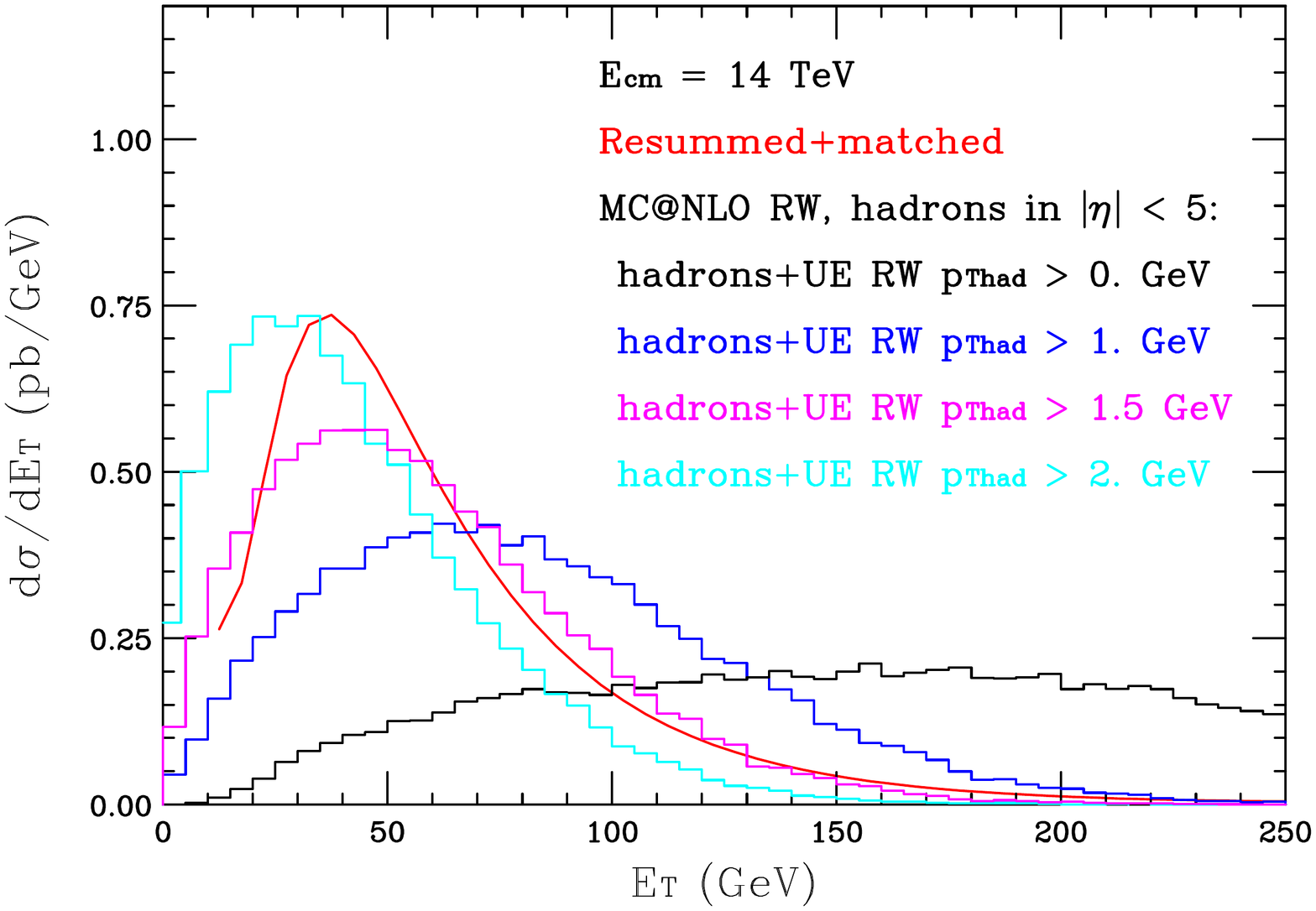}
  \caption{Hadron-level transverse-energy distribution in
  Higgs boson production at the LHC at 8 and 14 TeV, including the
  effect of the underlying event.
Red: resummed and matched to NLO, for comparison. Reweighted \aMC+\HW\
events with hadron maximum pseudorapidity $\eta^c = 5$:
Black: $p_{T}^c = 0$~GeV, Blue: $p_{T}^c = 1.0$~GeV, Magenta: $p_{T}^c
= 1.5$~GeV, Cyan: $p_{T}^c =2.0$~GeV. 
}
  \label{fig:mcnlo_hadronue_eta_pt}
\end{figure}

We present results using the default parameters present in
\HW\ version 2.6.3 for the underlying event model. We note that
these were tuned to a variety of experimental data using the MRST LO** PDF set~\cite{Sherstnev:2007nd}
instead of the MSTW2008 NLO set~\cite{Martin:2009iq} used here for the hard process generated using
\aMC.\footnote{MRST LO**, the default PDF set for LO processes in
  \HW, is called `MRSTMCal', with set number 20651 in the LHAPDF
  database~\cite{Whalley:2005nh}.}  In Fig.~\ref{fig:mcnlo_hadronue_eta_pt}
we show the $E_T$ distribution including the UE, with hadrons of a
maximum pseudorapidity $\eta^c = 5$, compared against the
analytical result, which we have shown matches well the hadron-level
$\eta^c = 5$ distribution without UE (Fig.~\ref{fig:mcnlo_hadron}).
In practice, particles cannot be detected at transverse momenta down
to zero, and therefore we show the
effect of applying transverse-momentum cuts on the hadrons: $p_{T}^c =
1.0, 1.5, 2.0$~GeV.  When $p_T^c=1.5$ GeV the peak in $E_T$ is moved
back to approximately the value of the parton-level prediction, but
the distribution itself is still somewhat broader.

We have also investigated the impact of the underlying
event using different PDFs and different, reasonable, model
parameters. We found that, with reasonably-tuned
values for the underlying event model parameters, the change of PDF
sets does not induce any significant changes to the distributions. 

To conclude, one can reproduce the $E_T$ distribution
with the effect of the UE and detector geometry effects by reweighting
the parton-level Monte Carlo events to match the analytical prediction
of the $E_T$ due to ISR and subsequently enabling the hadronization
and underlying event models of the generator. The description of the
underlying event is robust against changes of tune parameters as well
as PDF sets. However, in the presence of the underlying event the
$E_T$ distribution is highly sensitive to the minimum hadron transverse
momentum, $p_T^c$.

\section{Conclusions}\label{sec:conc}
We have presented the first detailed predictions of the
transverse-energy distribution in Higgs boson production at the LHC
($\sqrt{s}=8$ and $14$ TeV) for $m_H=126$ GeV.
Our calculation includes the resummation of the large logarithmic
terms at small $E_T$ up to (almost) NNLL accuracy, matched to the fixed-order
NLO result in a way that limits the impact of the resummation in the
intermediate and large-$E_T$ regions.

Our main result for the resummation is Eq.~(\ref{eq:resFnew}), with
component expressions (\ref{eq:RgRI}), (\ref{eq:gs}) and
(\ref{eq:Fal}).  For the matching we have Eq.~(\ref{eq:dsigmas}) with
(\ref{eq:RgNLO}), (\ref{eq:ETNLO}) and (\ref{eq:GH}).  The resulting
predictions are shown in Fig.~\ref{fig:sumEThigg126LHC}.  The
effect of resummation, compared to the pure NLO result, is large over
the whole range of $E_T$.  The purely resummed distribution peaks at
around 35 GeV and falls to unphysical negative values at small and
large $E_T$.  This behaviour is rectified by matching, which also
provides the NNLO normalization, without shifting the peak
significantly.  The uncertainty in the prediction, as assessed by the
customary factor-of-two variation of the renormalization scale, is
of the order of $\pm 10\%$.  However, the sensitivity
to unknown terms beyond NLO, in particular the NNLO coefficient
function $C^{(2)}_g$, is considerable, suggesting a larger
uncertainty. The possible impacts of  $C_g^{(2)}$ and the neglected
NNLL coefficient $A_g^{(3)}$ were illustrated in
Figs.~\ref{fig:sumEThigg126LHCC2} and \ref{fig:sumEThigg126LHCA3}
respectively.

The resummed and matched predictions refer only to the perturbative
hard-scattering component of Higgs production.  In real events there
are the non-perturbative effects of hadronization and the underlying
event.  We made a Monte Carlo study of these effects using \aMC\
interfaced to \HW\ or \PY, which provide a state-of-the-art simulation
of complete LHC final states.  The simulated events were reweighted at
the parton level to reproduce the analytically resummed and matched
$E_T$ distribution.  The effect of hadronization was to shift the peak
of the distribution upwards, to around 50 GeV, if all produced hadrons
were included.   However, this effect was practically eliminated by a
pseudorapidity cut $|\eta_{\rm had}|<5$.  The effect of the underlying event was
much greater, even in the presence of the pseudorapidity cut,  the
$E_T$ distribution becoming much broader, as was found in
Ref.~\cite{Papaefstathiou:2010bw}. This effect is due to soft hadrons
in the underlying event; a cut on hadron transverse momenta $p_{T{\rm had}}>1.5$
GeV restored the $E_T$ peak to around 30-40 GeV, although with a
distribution still somewhat broader than the parton-level prediction.

Measurements of differential distributions of the Higgs
boson at the LHC are starting to appear \cite{TheATLAScollaboration:2013eia}.
We look forward to measurement of the transverse-energy distribution and to
comparisons with our theoretical predictions.

\section*{Acknowledgements}
AP thanks Paolo Torrielli and Stefan Prestel for help in using the \aMC\ and \PY\
packages respectively and acknowledges support by the Swiss National
Science Foundation under contracts 200020- 138206 and 200020-141360/1.
AP and BW also acknowledge MCnetITN FP7 Marie Curie Initial Training Network
PITN-GA-2012-315877.
JMS is funded by a Royal Society University Research Fellowship.
BRW acknowledges the partial support of a Leverhulme Trust
Emeritus Fellowship and thanks
the Pauli Center for Theoretical Studies, Zurich, and the
Kavli IMPU, University of Tokyo, for hospitality and support during
parts of this work. The Kavli IPMU is supported by World Premier
International Research Center Initiative (WPI Initiative), MEXT, Japan.
\appendix

\section{Proof of an identity}\label{app:proof}
To prove Eq.~(\ref{eq:FZG}), we will show in general that
\beq
\left.f\left(\frac d{du}\right)\frac{Z(u)}u
\left[{\rm e}^{\lambda u}-1\right]\right|_{u=0}=
\left.Z\left(\frac d{d\lambda}\right)f\left(\frac d{du}\right)\frac 1u
\left[{\rm e}^{\lambda u}-1\right]\right|_{u=0}\,,
\eeq
where $f$ and $Z$ have power series expansions,
\beq
f(x)=\sum_\ell f_\ell\,x^\ell\,,\;\;\;
Z(u)=\sum_m Z_m\,u^m\,,
\eeq
which holds for the functions in Eq.~(\ref{eq:FZG}).  Now, beginning with the
left-hand side, 
\beq
\frac 1u\left[{\rm e}^{\lambda u}-1\right]
=\sum_{n=0}^\infty \frac{\lambda^{n+1}}{(n+1)!} u^n\,,
\eeq
so we can write
\beeq
\left.f\left(\frac d{du}\right)\frac{Z(u)}u \left[{\rm e}^{\lambda
      u}-1\right]\right|_{u=0}
&=&\sum_{\ell m n} \left. f_\ell\, Z_m\frac{d^\ell}{du^\ell} 
\frac{\lambda^{n+1}}{(n+1)!} u^{m+n}\right|_{u=0}\nn\\
&=&\sum_{\ell m n} f_\ell\, Z_m
\lambda^{n+1}\frac{(m+n)!}{(n+1)!}\,\delta_{\ell-m-n}\nn\\
&=&\sum_{\ell m} f_\ell\, Z_m
\lambda^{\ell-m+1}\frac{\ell!}{(\ell-m+1)!}\\
&=&\sum_{\ell m} f_\ell\, Z_m\frac{d^m}{d\lambda^m}
\frac{\lambda^{\ell+1}}{(\ell+1)}\nn\\
&=&\left.Z\left(\frac d{d\lambda}\right)f\left(\frac d{du}\right)\frac 1u
\left[{\rm e}^{\lambda u}-1\right]\right|_{u=0}\,.\nn
\eeeq

\section{Dispersion relations}\label{app:disp}
The fact that $d\sigma/dE_T\equiv F(E_T)$ has to vanish for $E_T<0$
implies that its Fourier transform $G(\tau)$
must satisfy dispersion relations
analogous to those in the frequency domain that follow from
causality.  Note first that if we write
\beq
F(E_T) = \Theta(E_T) f(E_T)\,,
\eeq
where $\Theta$ is the Heaviside step-function,
then $f(E_T)$ can be chosen to be either an odd or an even function.
We choose $f$ even, and then its Fourier transform $g(\tau)$ is purely real.
Now the Fourier transform of a product is a convolution, so
\beq
G(\tau) = \frac 1{2\pi} \int_{-\infty}^{+\infty} d\tau'\,h(\tau-\tau')
g(\tau')
\eeq
where $h$ is the Fourier transform of $\Theta$:
\beq
h(\tau)=\pi\,\delta(\tau) + P\frac i{\tau}\,,
\eeq
$P$ indicating the principal value.  Thus
\beq
G(\tau) = \frac 12 g(\tau) + \frac i{2\pi} P\int_{-\infty}^{+\infty}
d\tau'\,\frac{g(\tau')}{\tau-\tau'}\,.
\eeq
Now writing $G=G_R+iG_I$, recalling that $g(\tau)$ is real
and equating real parts we see that
\beq
g(\tau) = 2G_R(\tau)\,.
\eeq
Furthermore $G_I$ is not an independent function: it must satisfy the
dispersion relation
\beq
G_I(\tau) = \frac 1{\pi} P\int_{-\infty}^{+\infty}
d\tau'\,\frac{G_R(\tau')}{\tau-\tau'}\,.
\eeq
Notice that it follows that $G_R(\tau)$ must be an even function while
$G_I(\tau)$ must be odd, i.e.\ $G(-\tau)=G^*(\tau)$.  Altogether, we have 
\beq
G(\tau) = G_R(\tau) + \frac i{\pi} P\int_{-\infty}^{+\infty}
d\tau'\,\frac{G_R(\tau')}{\tau-\tau'}\,.
\eeq
Thus
\beq
F(E_T) = \frac 1{2\pi}\int_{-\infty}^{+\infty}d\tau\,{\rm
  e}^{-iE_T\tau}\left[ G_R(\tau) + \frac i{\pi} P\int_{-\infty}^{+\infty}
d\tau'\,\frac{G_R(\tau')}{\tau-\tau'}\right]\,.
\eeq
Assuming that the order of integration can be exchanged, the second
term involves
\beq
I(\tau')\equiv \frac i{\pi}P\int_{-\infty}^{+\infty}d\tau\,\frac{{\rm
  e}^{-iE_T\tau}}{\tau-\tau'}\,.
\eeq
The principal value implies the average of integrations along contours
above and below the pole at $\tau=\tau'$.  The contour can be closed
in the lower half-plane, where the exponential vanishes at infinity
since $E_T>0$.  Thus
\beq
I(\tau') = {\rm e}^{-iE_T\tau'}
\eeq
and, relabelling $\tau'$ as $\tau$ in the second term, we see that the
two
terms are equal and
\beq
F(E_T) = \frac 1{\pi}\int_{-\infty}^{+\infty}d\tau\,{\rm
  e}^{-iE_T\tau} G_R(\tau) \,.
\eeq
Thus we can simply replace the full Fourier transform $G$ by twice its
real part.  Furthermore, since $G_R$ is an even function, it then
follows immediately that
\beq
\int_0^\infty F(E_T)\,dE_T = G_R(0)\,.
\eeq
In the notation of Eq.~(\ref{eq:resFnew}), we have
\beeq
G_R(\tau)= {\rm e}^{-F_g^{(R)}(Q,\tau)} &\Bigl[&
 R_g^{(R)}(s;Q,\tau)\cos\{F_g^{(I)}(Q,\tau)\}\nn\\
&-&\,R_g^{(I)}(s;Q,\tau)\sin\{F_g^{(I)}(Q,\tau)\}\;\Bigr]
\,\sigma_{gg}^H(Q,\as(Q))
\eeeq
and, by virtue of the shift (\ref{eq:ltotl}),
$F_g^{(R)}(Q,0)=F_g^{(I)}(Q,0)=0$, so we obtain Eq.~(\ref{eq:unit}).

\section{Comparison with transverse-momentum
resummation}\label{app:ptcomp}
For the Higgs transverse momentum $q_T$, instead of integrals of the
form (\ref{eq:Ip}) we have\footnote{Here we ignore the shift in the argument of the
logarithm, which gives only power corrections.}
\beq\label{eq:bIp}
\bI_p(q_T,Q) = q_T\int_0^{\infty} db\,b J_0(bq_T)
\ln^p\left(\frac{bQ}{b_0}\right)\;,
\eeq
where $b_0=2\exp(-\gE)$.  These integrals may be evaluated from
\beq\label{eq:bIpIu}
\bI_p(q_T,Q) = \frac{d^p}{du^p}\bI(q_T,Q;u)|_{u=0}
\eeq
where
\beeq\label{eq:bIu}
\bI(q_T,Q) &=& q_T\int_0^{\infty} db\,b J_0(bq_T)
\left(\frac{bQ}{b_0}\right)^u\nn\\
&=&-\frac{2{\rm e}^{u\gE}}{\pi q_T}\left(\frac Q{q_T}\right)^u\sin\left(\frac{\pi
    u}2\right)\Gamma^2\left(1+\frac u2\right)\;,
\eeeq
which can be written as
\beeq
\bI(q_T,Q)&=&-\frac{u{\rm
    e}^{u\gE}}{q_T}\left(\frac Q{q_T}\right)^u\frac{\Gamma(1+u/2)}{\Gamma(1-u/2)}\nn\\
&=& -\frac u{q_T}\left(\frac
  Q{q_T}\right)^u\exp\left[-2\sum_{k=1}^\infty \frac{\zeta_{2k+1}}{2k+1}\left(
\frac u2\right)^{2k+1}\right]\;.
\eeeq
This gives instead of Eq.~(\ref{eq:I12})
\beeq\label{eq:bI12}
\bI_1(q_T,Q) &=& -\frac 1{q_T}\;,\quad
\bI_2(q_T,Q) = -\frac 2{q_T}\ln\left(\frac Q{q_T}\right)\nn\\
\bI_3(q_T,Q) &=& -\frac3{q_T}\ln^2\left(\frac Q{q_T}\right)\nn\\
\bI_4(q_T,Q) &=& -\frac4{q_T}\left[\ln^3\left(\frac Q{q_T}\right)-\frac 12\z3\right]\;.
\eeeq
Therefore at small $q_T>0$ we expect
\beq\label{eq:qTNLO}
\left[\frac{q_T}{\sigma_0^H}\frac{d\sigma_H}{dq_T}\right]_{\rm NLO}\sim
  \bas(\bG_0 R_0+\bG'_1 R'_1)+\bas^2(\bH_0 R_0+\bH_1 R_1+\bH'_1 R'_1
+\bH'_2 R'_2+\bH''_2 R''_2)
\eeq
where, writing $L=\ln(Q/q_T)$,
\beeq\label{bGbH}
\bG_0&=&\bH_1=4A_g^{(1)}L+2B_g^{(1)}\;,\;\;\;
\bG'_1=-1\;,\nn\\
\bH_0&=&4L\left[A_g^{(2)}+\beta_0 B_g^{(1)} -\beta_0 A_g^{(1)}
\ln\left(\frac{Q^2}{\mu_R^2}\right)\right] 
+8\beta_0 A_g^{(1)}L^2\nn\\
&+&2\left[\overline B_g^{(2)}-\beta_0 B_g^{(1)}
\ln\left(\frac{Q^2}{\mu_R^2}\right)\right]
-8(A_g^{(1)})^2\left(L^3-\frac 12\z3\right)\nn\\
&-&12A_g^{(1)}B_g^{(1)}L^2-4(B_g^{(1)})^2 L\;,\nn\\
\bH'_1&=&6A_g^{(1)}L^2+4B_g^{(1)} L\;,\nn\\
\bH'_2 &=&-1\;,\;\;\; \bH''_2=-2L\;.
\eeeq
Here we have allowed for the possibility that the coefficient
$\overline B_g^{(2)}$ for $q_T$ may be different from $B_g^{(2)}$ for $E_T$.
Comparing with Eqs.~(\ref{eq:ETNLO}) and (\ref{eq:GH}), we see that
the NLO $E_T$ and $q_T$ distributions at the point $q_T=E_T$ differ by 
\beq\label{eq:EtPtdif}
\frac{d\sigma_H}{dE_T} - \left.\frac{d\sigma_H}{dq_T}\right|_{q_T=E_T} \sim \;\;
\bas^2\,\frac{\sigma_B^H}{E_T
}\left[H_0-\bH_0+(H'_1-\bH'_1)\frac{R'_1}{R_0}\right]
\eeq
where $\sigma_B^H=\sigma_0^H R_0$ is the Born cross section and
\beeq
H_0-\bH_0 &=& \frac 43\pi^2 A_g^{(1)}\left(2A_g^{(1)}L +
  B_g^{(1)}\right)
+2\left(B_g^{(2)}-\overline B_g^{(2)}\right)+12\z3\left(A_g^{(1)}\right)^2\,,\nn\\
H'_1-\bH'_1 &=& -\frac 23\pi^2 A_g^{(1)}\,.
\eeeq

\section{Alternative Monte Carlo results}\label{app:pythia}
\begin{figure}[!htb]
    \includegraphics[width=0.38\linewidth,
    angle=90]{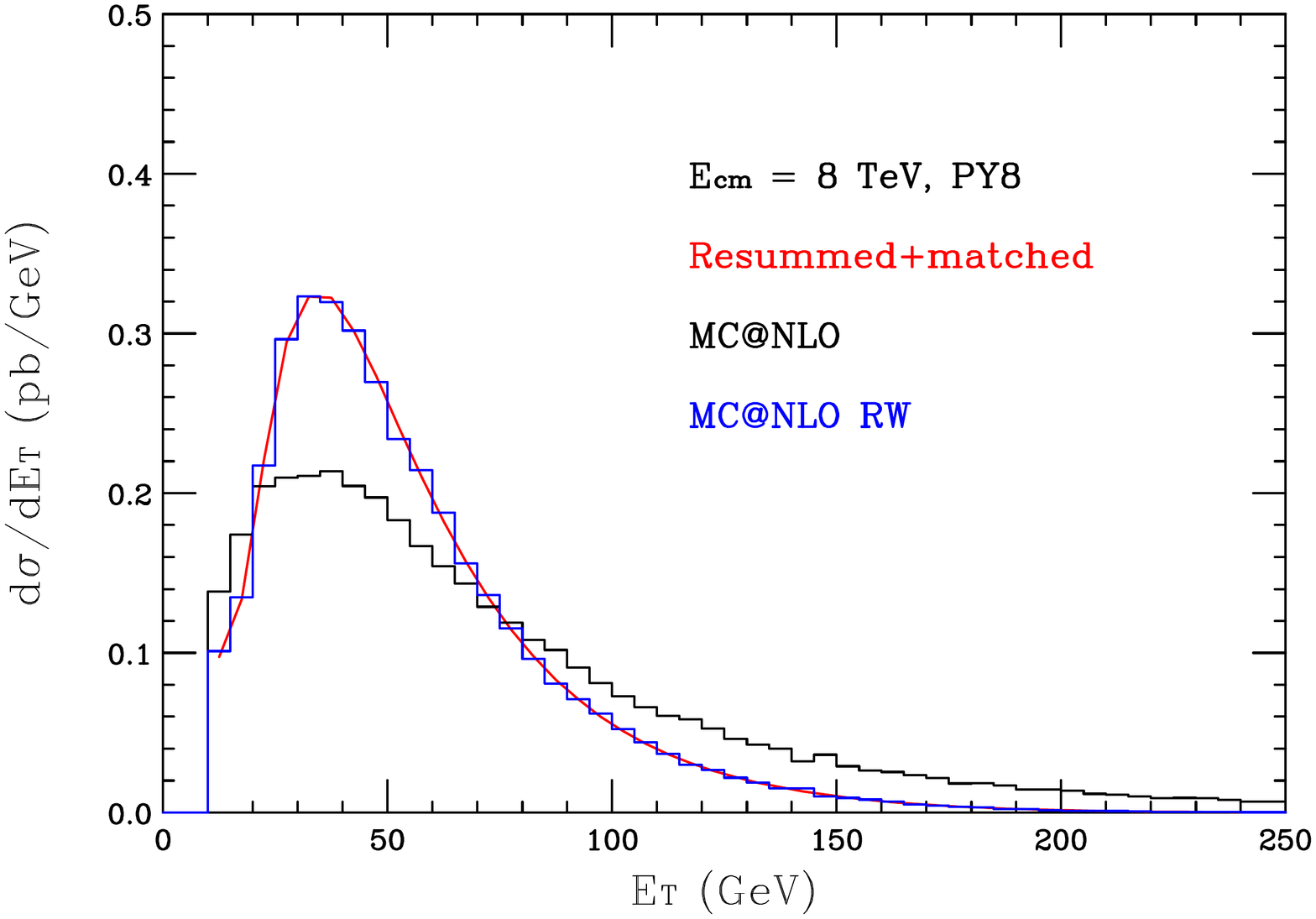}
    \includegraphics[width=0.38\linewidth, angle=90]{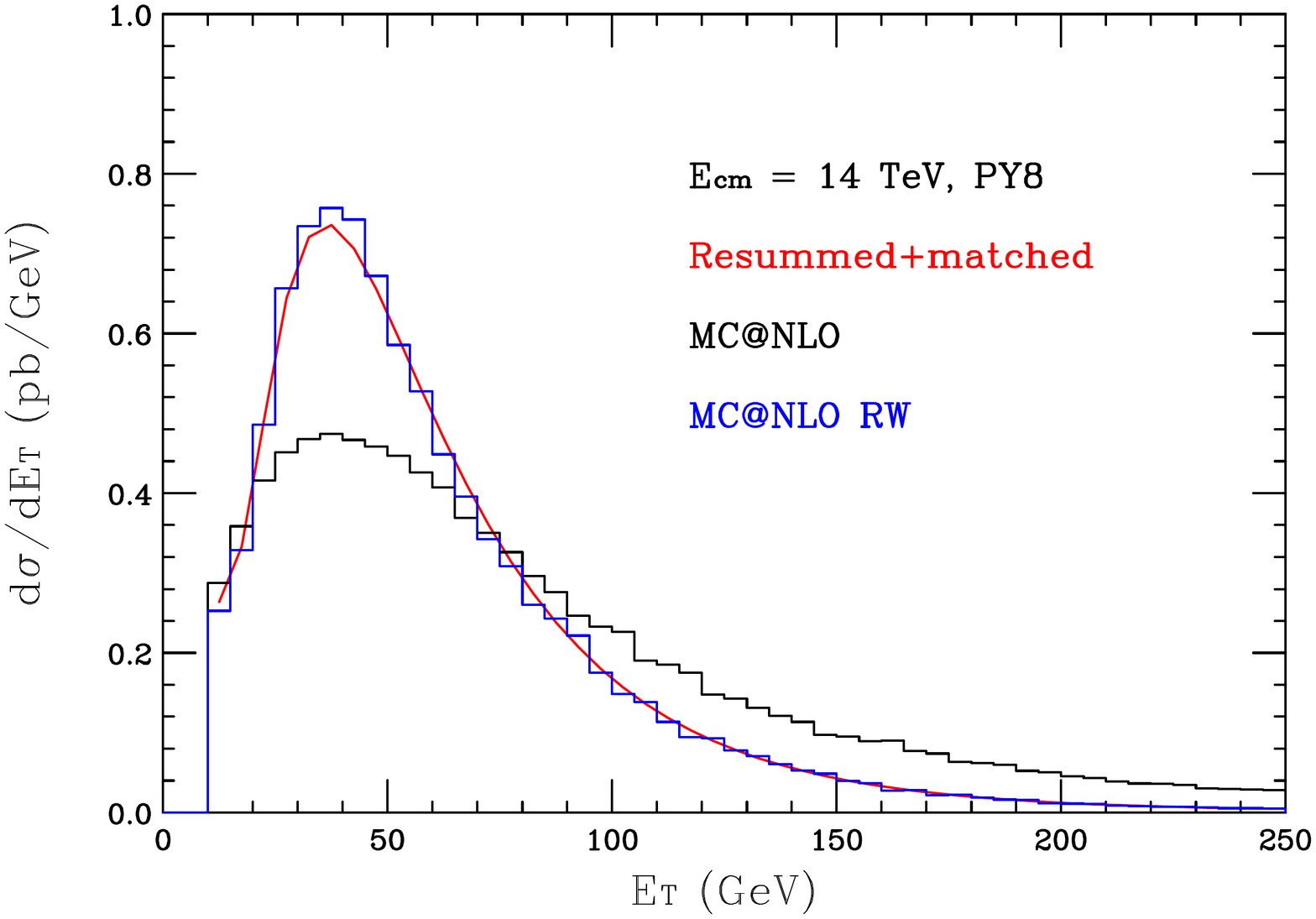}
  \caption{Parton-level transverse-energy distribution in
  Higgs boson production at the LHC at 8 and 14 TeV.
Red: resummed and matched to NLO.
Black: \aMC+\PY\ before reweighting.
Blue: \aMC+\PY\ after reweighting.}
  \label{fig:mcnlopy8}
\end{figure}

For purposes of comparison with alternative descriptions of the parton
shower, hadronization and the underlying event, we provide here results
equivalent to those obtained using \HW\ in Sec.~\ref{sec:MC}, using
the \PY\ event
generator~\cite{Sjostrand:2006za,Sjostrand:2007gs}. We use the default parameters appearing in
\PY\, version 8.185, with the Higgs boson mass set to
126~GeV. Figure~\ref{fig:mcnlopy8} is equivalent to Fig.~\ref{fig:mcnlo} for \HW\ and demonstrates that the reweighting procedure reproduces the analytical resummed and matched result. 

\begin{figure}[!htb]
    \includegraphics[width=0.38\linewidth,
    angle=90]{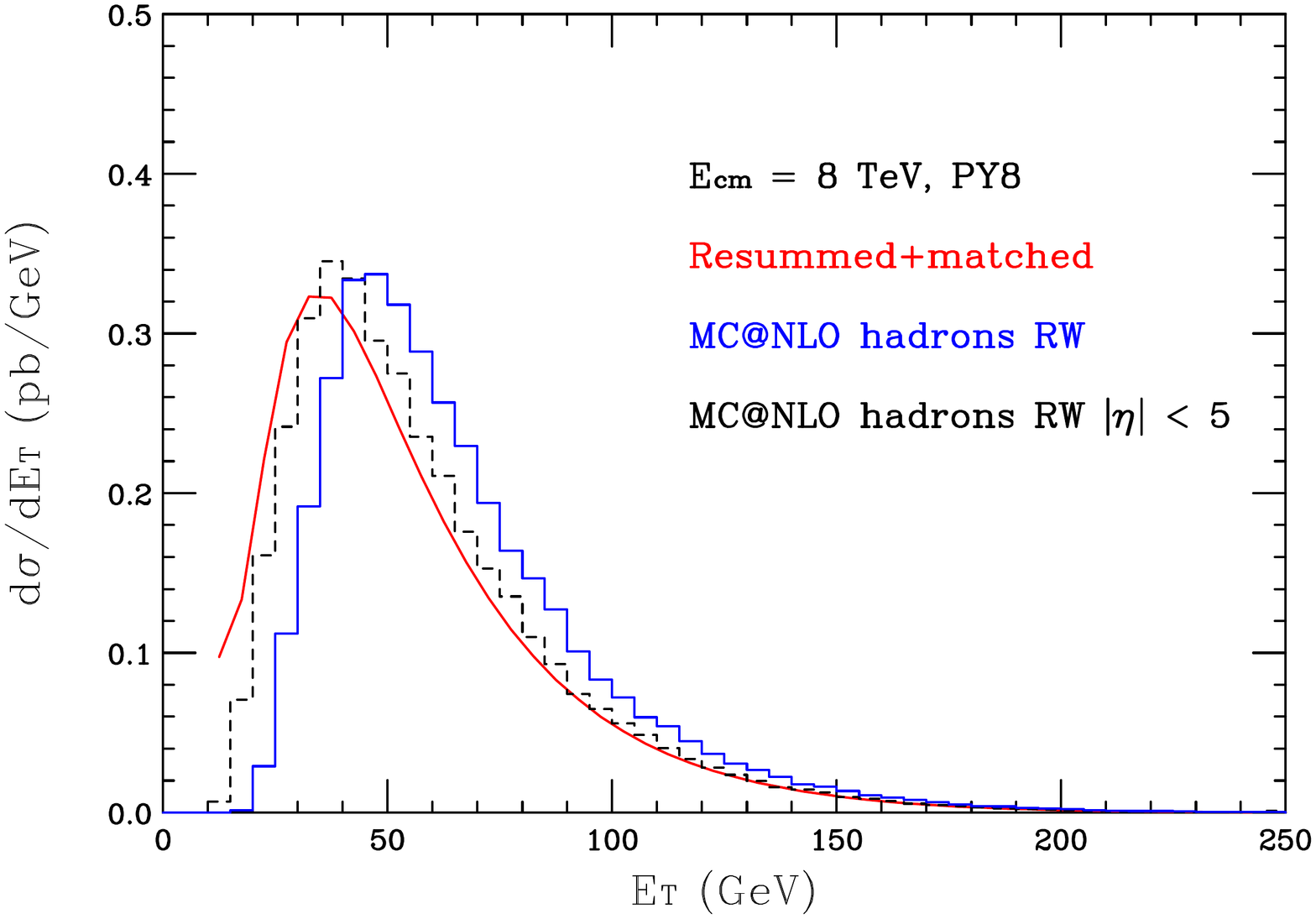}
    \includegraphics[width=0.38\linewidth, angle=90]{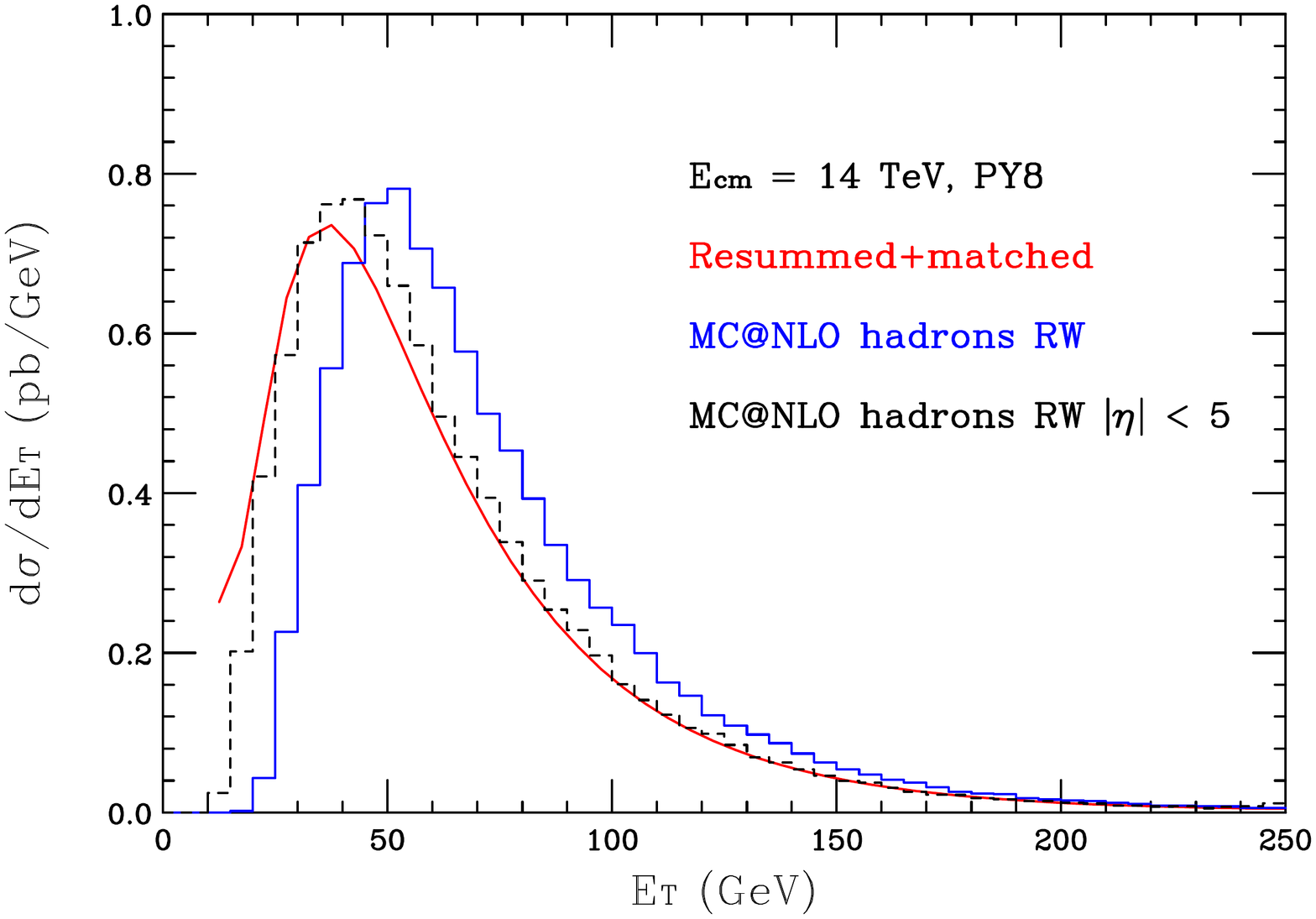}
  \caption{Hadron-level transverse-energy distribution in
  Higgs boson production at the LHC at 8 and 14 TeV.
Red: resummed and matched to NLO.
Blue: \aMC+\PY.
Black dashes: \aMC+\PY\, particles
restricted to lie within pseudorapidity $|\eta| < 5$.
}
  \label{fig:mcnlo_hadron_py}
\end{figure}

In Fig.~\ref{fig:mcnlo_hadron_py} we show the effect of hadronization on the parton-level $E_T$ distribution. Comparing to Fig.~\ref{fig:mcnlo_hadron}, it can be observed that the effect is of similar magnitude and the compensation obtained by applying a cut of $|\eta|<5$ persists in \PY.

\begin{figure}[!htb]
    \includegraphics[width=0.38\linewidth,angle=90]{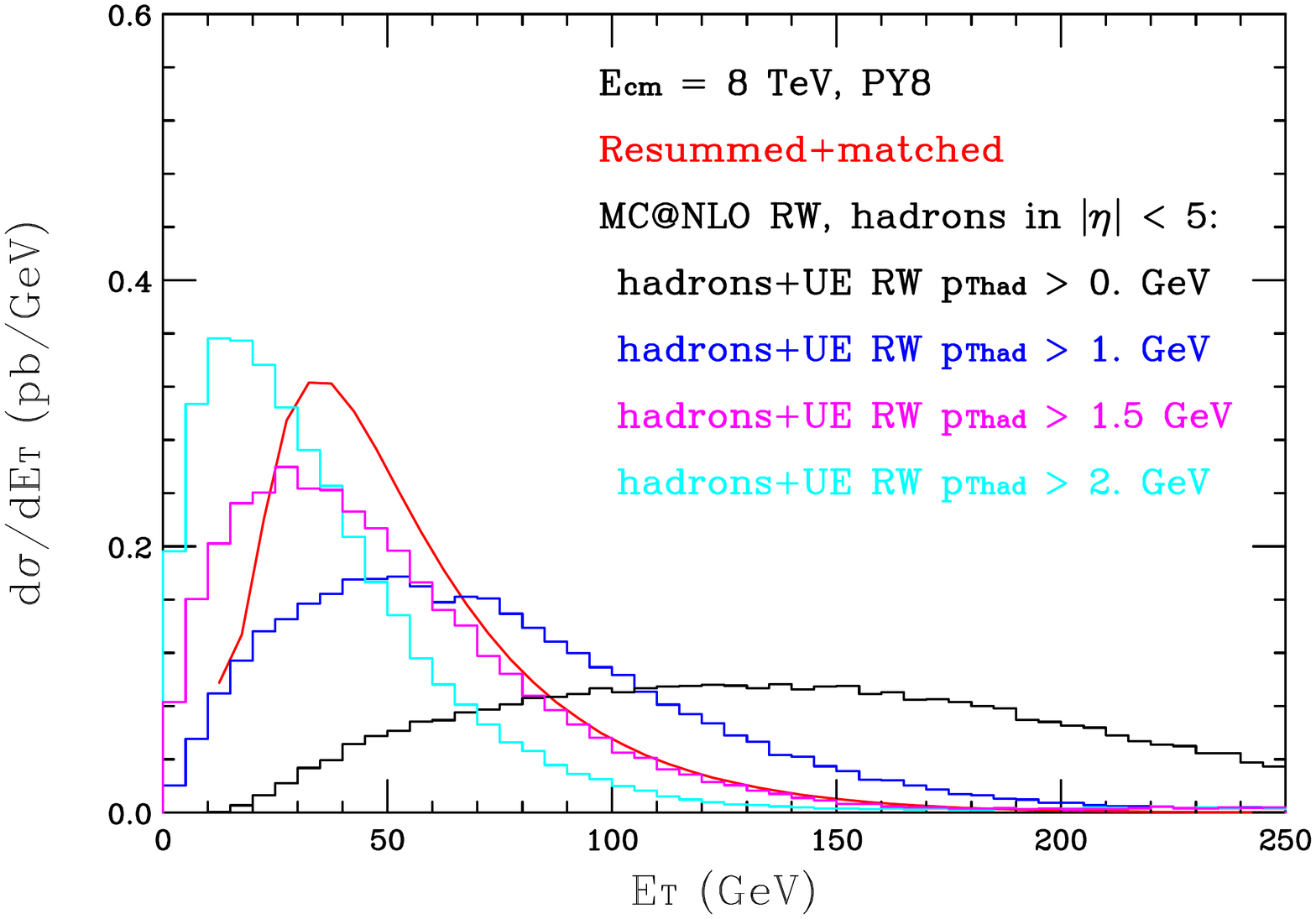}
    \includegraphics[width=0.38\linewidth, angle=90]{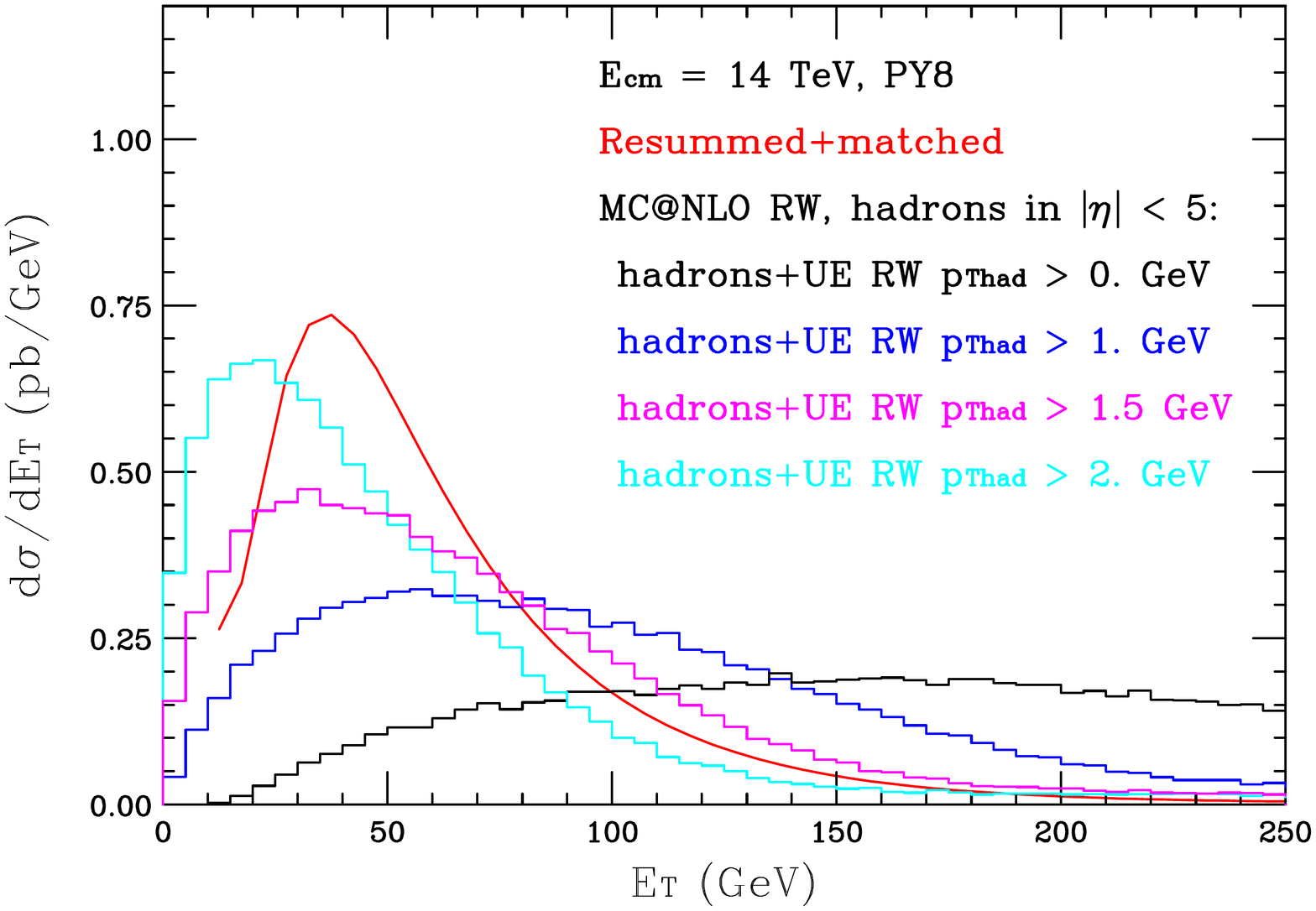}
  \caption{Hadron-level transverse-energy distribution in
  Higgs boson production at the LHC at 8 and 14 TeV, including the
  effect of the underlying event.
Red: resummed and matched to NLO.
\aMC+\PY\ events with hadron maximum pseudorapidity $\eta^c = 5$:
Black: $p_{T}^c = 0$~GeV, Blue: $p_{T}^c = 1.0$~GeV, Magenta: $p_{T}^c
= 1.5$~GeV, Cyan: $p_{T}^c =2.0$~GeV. 
}
  \label{fig:mcnlo_hadronue_eta_pt_pythia}
\end{figure}

The effect of the underlying event model present in \PY\
is shown in Fig.~\ref{fig:mcnlo_hadronue_eta_pt_pythia}. Evidently,
the effect is qualitatively similar to what was shown in
Fig.~\ref{fig:mcnlo_hadronue_eta_pt} for \HW. Moreover, the effect of imposing a minimum transverse momentum on the contributing hadrons is also identical to that observed in \HW.


\end{document}